\documentclass[conference,compsoc]{IEEEtran}

\IEEEoverridecommandlockouts

\usepackage{cite}

\usepackage{amsmath,amssymb,amsfonts}
\usepackage{graphicx}
\usepackage{booktabs}
\usepackage{diagbox}
\usepackage{multirow}
\usepackage{tabularx}
\usepackage{enumitem}
\usepackage{xspace}
\usepackage{bm}
\usepackage{pifont}
\usepackage{xurl}
\usepackage[hidelinks]{hyperref}
\usepackage{subcaption}
\usepackage{adjustbox}
\usepackage{colortbl}
\usepackage[table]{xcolor}
\usepackage{tikz}
\usepackage{pgfplots}
\usepgfplotslibrary{polar,groupplots}
\usetikzlibrary{positioning,shapes.geometric,arrows.meta,calc,patterns}
\usepackage{pgfplotstable}

\pgfplotsset{compat=newest}

\usepackage[ruled,vlined,linesnumbered]{algorithm2e}

\pgfplotsset{
  every axis/.append style={
    ymajorgrids=true,
    grid style={dotted},
    tick label style={font=\footnotesize},
    label style={font=\footnotesize},
    title style={font=\footnotesize},
  }
}

\newcommand{\NS}{\textit{NS}\xspace}
\newcommand{\NSes}{\textit{NSes}\xspace}
\newcommand{\NSAttack}{\textit{NS attack}\xspace}
\newcommand{\NegativeShadow}{\textit{Negative Shadow (NS)}\xspace}
\newcommand{\NegativeShadows}{\textit{Negative Shadows (NSes)}\xspace}
\newcommand{\NegativeShadowAttack}{\textbf{\textit{Negative Shadow (NS) Attack}}\xspace}

\definecolor{mygreen}{RGB}{0,128,0}

\newcommand{\xmark}{\ding{55}}
\newcommand{\cmark}{\ding{51}}

\begin{document}

\title{On the Vulnerability of Lane Detectors to Physical Shadow Attack\vspace{-1.5em}}

\author{
\IEEEauthorblockN{
Pedram MohajerAnsari\IEEEauthorrefmark{1},
Amir Salarpour\IEEEauthorrefmark{1},
Jan de Voor\IEEEauthorrefmark{1},
Arkajyoti Mitra\IEEEauthorrefmark{2},\\
Alkim Domeke\IEEEauthorrefmark{1},
Habeeb Olufowobi\IEEEauthorrefmark{2},
Mohammad Hamad\IEEEauthorrefmark{3},
Mert D. Pesé\IEEEauthorrefmark{1}
}

\IEEEauthorblockA{
\IEEEauthorrefmark{1}Clemson University
\qquad
\IEEEauthorrefmark{2}University of Texas at Arlington
\qquad
\IEEEauthorrefmark{3}Technical University of Munich
}
}

% \author{
% \IEEEauthorblockN{
% Pedram Mohajeransari\IEEEauthorrefmark{1},
% Amir Salarpour\IEEEauthorrefmark{1},
% Jan de Voor\IEEEauthorrefmark{1},
% Arkajyoti Mitra\IEEEauthorrefmark{2},\\
% Alkim Domeke\IEEEauthorrefmark{1},
% Habeeb Olufowobi\IEEEauthorrefmark{2},
% Mohammad Hamad\IEEEauthorrefmark{3},
% Mert D. Pesé\IEEEauthorrefmark{1}
% }

% \IEEEauthorblockA{
% \IEEEauthorrefmark{1}Clemson University, Clemson, SC, USA; 
% \{pmohaje, asalarp, jdevoor, adomeke, mpese\}@clemson.edu
% }

% \IEEEauthorblockA{
% \IEEEauthorrefmark{2}University of Texas at Arlington, Arlington, TX, USA; 
% \{arkajyoti.mitra, habeeb.olufowobi\}@uta.edu
% }

% \IEEEauthorblockA{
% \IEEEauthorrefmark{3}Technical University of Munich, Munich, Germany; 
% mohammad.hamad@tum.de
% }
% }

\IEEEpubid{\parbox{\columnwidth}{\footnotesize \copyright~2026 IEEE. Personal use of this material is permitted. Permission from IEEE must be obtained for all other uses, in any current or future media, including reprinting/republishing this material for advertising or promotional purposes, creating new collective works, for resale or redistribution to servers or lists, or reuse of any copyrighted component of this work in other works.}\hspace{\columnsep}\makebox[\columnwidth]{}}
\maketitle

\begin{abstract}
Automated Lane Centering (ALC) systems are critical components of Level 2 autonomous vehicles, yet their reliance on Deep Neural Networks (DNNs) for lane detection introduces security vulnerabilities. This work introduces the Negative Shadow (NS) Attack, a novel physical-world adversarial technique that exploits intensity bias in shadow-removal preprocessing, where bright patches within shadowed regions are misclassified as genuine lane markings. Unlike previous attacks requiring illegal road markings, the NS attack is passive and stealthy, utilizing sunlight filtered through shaped occluders placed on private property to cast patterns onto the roadway. We utilize a genetic algorithm to optimize the geometry of these shadows against three state-of-the-art lane detection systems: CLRerNet, HybridNets, and TwinLiteNet. The attack is evaluated across three hazard scenarios—left-turn deviation, head-on collision, and off-road deviation—using software-in-the-loop simulations with CARLA-OpenPilot, which demonstrated violation rates of up to 100\%. Further validation is provided through a miniature road testbed and a real-world deployment where the attack successfully caused a Comma 3X-equipped vehicle to depart its lane. Additionally, a human-subject study indicates high stealth, with 83.6\% of participants failing to detect the adversarial patterns, and an assessment of input-preprocessing defenses shows they offer only limited protection against this threat.
\end{abstract}


\section{Introduction}
\label{sec:1-Introduction}
Automated Lane Centering (ALC) systems play a critical role in modern Level~2 autonomous vehicles (AVs) by maintaining lane position during driving, and have been widely adopted in commercial platforms such as Tesla Autopilot~\cite{tesla2023autopilot}. These systems rely heavily on lane detection (LD) algorithms that process camera input to identify lane boundaries~\cite{becker2018functional}. Recent work increasingly uses deep neural networks (DNNs) for LD, improving accuracy in challenging environmental conditions~\cite{pan2018spatial, padmaraja2023lane}. % and Mercedes-Benz Drive Pilot~\cite{mercedesbenz2023drivepilot}

However, the adoption of DNN-based LD systems introduces an attack surface. These models are known to be susceptible to adversarial examples (AEs), i.e., small input perturbations that can cause misclassification~\cite{goodfellow2014explaining, kurakin2018adversarial}. Several studies have extended this vulnerability into the physical world, including the Black-Strip~\cite{boloor2019simple}, Dirty Road Patch (DRP)~\cite{sato2021dirty}, and RAP-ALC~\cite{caoremote} attacks, which demonstrate the feasibility of deceiving ALC systems. Yet, these attacks face limitations: DRP and Black-Strip require illegal alterations to public infrastructure, while RAP-ALC lacks stealth, as it requires a patch attached to the attacker’s car.

\begin{figure}[t] %tbp
    \centering    
    \includegraphics[width=0.9\columnwidth]{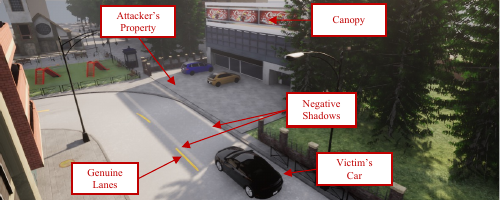}
    \caption{An attacker uses a canopy with two holes to cast bright patterns (\emph{Negative Shadow}) that mimic line markings, causing the victim vehicle to proceed straight rather than turn left, resulting in a collision with the playground ahead. The setup remains discreet and does not violate road laws.}
    \label{fig:teaser_picture}
\end{figure}

In this work, we propose a novel physical-world attack called the \NegativeShadowAttack, which alters AV perception using sunlight rather than physical markings. The core insight is that LD algorithms show an \emph{intensity bias} during shadow-removal preprocessing: bright patches within shadowed regions tend to be classified as non-shadows~\cite{zhu2021mitigating}. Our attack exploits this bias by passing sunlight through a shaped occluder, such as a canopy or fence with holes, positioned on private property. This creates a \emph{{\textbf{bright, line-like pattern}}} referred as a \emph{{\textbf{Negative Shadow (\NS)}}} on the road surface that AVs interpret as real lane markings. This attack is \emph{passive} and \emph{stealthy}, as it resembles incidental lighting, and \emph{legal}, avoiding any tampering with public infrastructure.

%\colorbox{black!5}

\autoref{fig:teaser_picture} shows a scenario: an AV misclassifies the bright stripes (each a \NegativeShadow, collectively \NegativeShadows) as lane markings, labels the region between them as drivable space, and drives straight into a playground instead of executing a left turn. Because the \NSes are created by sunlight and look natural on the road, the attack is hard for human drivers and regulators to detect visually.

To explore the efficacy of the \NSAttack, we design a genetic algorithm that searches over the \NS geometry (length ($L$), width ($W$), lateral offset ($D$), and angle ($\beta$)) to find configurations that are consistently misdetected as lane markings. The composite fitness function evaluates each candidate by the fraction of \NS pixels that LD models label as lane markings. We run this search against three state-of-the-art DNN-based LD systems: CLRerNet~\cite{honda2023clrernet}, HybridNets~\cite{vu2022hybridnets}, and TwinLiteNet~\cite{che2023twinlitenet} which we select for their high reported accuracy and to span diverse LD architectures. This evaluation lets us identify which \NS lengths, widths, offsets, and angles most reliably cause LD models to misdetect \NSes as lane markings.

We consider three scenarios: (i) a \textbf{left-turn deviation}, (ii) a \textbf{head-on collision}, and (iii) an \textbf{off-road deviation}, and evaluate the attack’s impact across four experimental setups: (1) \textbf{software-in-the-loop simulations} using CARLA-OpenPilot \cite{commaaiopenpilot2023}, covering all three scenarios and demonstrating up to 100\% violation rates in the left-turn and head-on cases; (2) a \textbf{miniature road testbed} that reproduces the head-on collision scenario on a scaled physical road with an off-vehicle Comma~3X; (3) a \textbf{real-world deployment} that reproduces the off-road deviation scenario, where the \NSAttack causes a Comma~3X-equipped vehicle~\cite{comma3x} to depart the lane and collide with a barrier; and (4) a \textbf{human-subject study} showing that 83.6\% of participants fail to detect the \NS pattern under simulated driving conditions. Finally, we assess several input-preprocessing defenses and find that they provide only limited protection against the \NSAttack. This paper makes the following contributions:
\begin{enumerate}[leftmargin=*, itemsep=0pt, topsep=0pt]
    \item We introduce the \NegativeShadowAttack, a novel, stealthy, and non-invasive physical-world adversarial attack that exploits intensity bias in lane detection shadow-removal preprocessing. Unlike prior attacks requiring illegal road modifications, the \NSAttack uses sunlight filtered through shaped occluders on private property to cast deceptive bright patterns that mimic lane markings.
    \item We develop a genetic algorithm to systematically search the \NS parameter space (length, width, lateral distance, and angle) and demonstrate that our attack reliably misleads three state-of-the-art DNN-based lane detection models (CLRerNet, HybridNets, and TwinLiteNet) across diverse architectures. Code and supplementary materials are available in our repository: \textcolor{blue}{\url{https://github.com/pedram-mohajer/NS-Attack}}.

    \item We comprehensively evaluate the \NSAttack across four experimental setups covering three safety-critical scenarios (left-turn deviation, head-on collision, and off-road deviation): (1) software-in-the-loop simulations achieving up to 100\% violation rates, (2) a miniature road testbed validating physical feasibility, (3) real-world deployment successfublly causing lane departure in a Comma~3X-equipped vehicle, and (4) human-subject study demonstrating 83.6\% missed detection rate, confirming high stealth.
    \item We assess existing input-preprocessing defenses including bit-depth reduction, Gaussian noise, median blur, and JPEG compression, as well as Kalman-filtered temporal tracking, finding limited protection from preprocessing, while temporal tracking prevents 58.5--68.0\% of \NS-induced false-lane acceptance.
\end{enumerate}

\section{Background}
\label{sec:2-Background}
\noindent \textbf{Lane Detection in Autonomous Driving.} LD algorithms are essential for navigation in AV systems as they detect and track line markings to guide the vehicle~\cite{waykole2021review, sheu2021driving}. These algorithms are broadly classified into two categories: feature-based and model-based~\cite{wang2010model}. Feature-based algorithms analyze the visual characteristics of images captured by vehicle cameras, relying on techniques such as edge detection (e.g., Sobel operator~\cite{sobel1968feldman}, Canny edge detector~\cite{canny1986computational}), texture analysis, and color segmentation~\cite{kang2003road, zheng2019lane}. Color-based thresholds are also frequently used to distinguish bright line markings from darker road backgrounds~\cite{yang2019improved}. In contrast, model-based algorithms use mathematical representations, such as polynomial regression or spline models, to estimate line geometries~\cite{dickmanns1992recursive, bertozzi1998gold}. These models fit detected line markings into predefined geometric frameworks, often using polynomial equations to approximate the lane curvature~\cite{wang2000lane, wang2004lane}. Geometric considerations, such as perspective transformations, further simplify lane detection by providing a bird's-eye view of the road~\cite{zakaria2023lane}. 

Despite advancements, LD algorithms still rely on common techniques, resulting in vulnerabilities shared across different implementations. A key flaw is \textit{intensity bias} in shadow detection, where over-reliance on brightness levels leads to misdetection, such as mistaking bright areas within shadows for line markings or incorrectly identifying dark regions as shadows~\cite{zhu2021mitigating, chen2022semi}. To mitigate this, shadow removal pre-processing is commonly used to reduce or eliminate shadows from sensor data~\cite{hoang2017road}. However, even with advanced methods such as feature decomposition~\cite{watson2016entropy} and reweighting schemes~\cite{maimon2002improving}, these vulnerabilities persist due to the reliance on such preprocessing.

\noindent \textbf{Shadow Removal in Lane Detection.}
Shadow removal algorithms aim to recover a shadow-free version of an image, denoted as $\hat{I}_{sf}$, from an input shadow image $I_s$. Modern approaches use a DNN-based shadow removal network $G$ with trainable parameters $\theta$ to learn this mapping from data:
\begin{equation}
    \hat{I}_{sf} = G(I_s; \theta).
\end{equation}

Many recent methods further exploit a shadow mask $M$, obtained either from annotations or from a learned shadow detector, to guide this process~\cite{liu2021shadow}. For example, ST-CGAN jointly learns a shadow mask and a shadow-free image in a stacked GAN architecture~\cite{wang2018stacked}, DHAN uses a mask-based attention loss for ghost-free removal~\cite{cun2020towards}, CNSNet explicitly conditions its normalization and transformer modules on the shadow mask~\cite{yu2022cnsnet}, and ShadowFormer takes a shadow image together with its mask as input to a transformer-based restoration network~\cite{guo2023shadowformer}. Even when the mask is not directly fed to the final generator (e.g., in LG-ShadowNet), it is often estimated internally or used during training to better localize shadow regions~\cite{hu2019mask}.

Although these methods are implemented as deep networks, their design is grounded in the classical Retinex theory of color vision~\cite{land1974retinex}, which models an image as the element-wise product of illumination $L$ and reflectance $R$. Under this view, a shadow-free image can be written as
\begin{equation}
    I_{sf} = L_{sf} \odot R,
\end{equation}
where $L_{sf}$ denotes the illumination field (lighting) in a shadow-free scene and $R$ is the scene reflectance (intrinsic surface color/texture).

\begin{table}[t]
\centering
\scriptsize
\setlength{\tabcolsep}{2pt}
\renewcommand{\arraystretch}{1.1}
\caption{Comparison of various perception (ALC) attacks based on stealthiness, practicality, and law violation}
\label{tab:attacks_comparison}
\begin{tabular}{l*{5}{c}}
\toprule
 & \textbf{Black-Strip} & \textbf{DRP} & \textbf{Crafted} & \textbf{RAP-ALC} & \cellcolor{green!10} \textbf{NSAttack} \\
 & \textbf{Attack}~\cite{boloor2019simple} & \textbf{Attack}~\cite{sato2021dirty} & \textbf{Perturb.}~\cite{jing2021too} & \cite{caoremote} & \cellcolor{green!10}\textbf{(Ours)} \\
\midrule
\textbf{Stealthy}       & \cmark & \cmark & \cmark & \xmark & \cellcolor{green!10}\cmark \\
\textbf{Practical}      & \cmark & \xmark & \xmark & \cmark & \cellcolor{green!10}\cmark \\
\textbf{Law Violation}  & Directed & Directed & Directed & Indirected & \cellcolor{green!10}Indirected \\
\bottomrule
\end{tabular}
\end{table}

% \begin{table}[t]
% \centering
% \footnotesize
% \setlength{\tabcolsep}{4pt}
% \renewcommand{\arraystretch}{1.15}
% \caption{Comparison of various perception (ALC) attacks based on stealthiness, practicality, and law violation}
% \label{tab:attacks_comparison}

% \begin{tabularx}{\columnwidth}{
%   >{\raggedright\arraybackslash}X
%   >{\centering\arraybackslash}p{0.7cm}
%   >{\centering\arraybackslash}p{0.7cm}
%   >{\centering\arraybackslash}p{2.0cm}
% }

% \toprule
% \multicolumn{4}{c}{\hspace{3cm}\textbf{Operational Parameters}} \\
% \cmidrule(lr){2-4}
% \textbf{Attacks} & \textbf{Stealthy} & \textbf{Practical} & \textbf{Law Violation} \\
% \midrule
% Black-Strip Attack~\cite{boloor2019simple} & \cmark & \cmark & \textbf{Directed} \\
% DRP-Attack~\cite{sato2021dirty}            & \cmark & \xmark & \textbf{Directed} \\
% Crafted-Perturbations Attack~\cite{jing2021too}     & \cmark & \xmark & \textbf{Directed} \\
% RAP-ALC~\cite{caoremote}                   & \xmark & \cmark & \textbf{Indirected} \\
% \rowcolor{green!10}\NegativeShadowAttack   & \cmark & \cmark & \textbf{Indirected} \\
% \bottomrule
% \end{tabularx}
% \end{table}

This represents the ideal shadow-free image obtained under perfect conditions, and helps disentangle how illumination and reflectance contribute to the appearance of shadows. Shadows arise from degraded illumination $L_s$ in part of the scene, while $R$ remains unchanged, leading to a shadowed image
\begin{equation}
    I_s = L_s \odot R = A \odot L_{sf} \odot R,
\end{equation}
where $A$ is a spatially varying degradation factor (typically $0 < A \leq 1$ in shadow regions) that attenuates the original illumination~\cite{qu2017deshadownet}. The goal of shadow removal is to estimate and compensate for this degradation, thereby restoring the original illumination and recovering the shadow-free image $\hat{I}_{sf}$. In the context of LD pipelines, such pre-processing is intended to suppress dark cast shadows that obscure lane markings, reducing false negatives due to shadowed lines. At the same time, dark shadow patterns that resemble lane markings (e.g., elongated dark stripes on the road surface) would also tend to be corrected by these methods, since they are explicitly designed to detect and mitigate shadow-like intensity variations before downstream lane detection.

\noindent \textbf{Automated Lane Centering.} A typical ALC system design operates in three steps~\cite{sato2021dirty}. First, the process begins in the perception layer, which processes outputs from LD algorithms, including techniques such as Eigenvalue Decomposition Regularized Analysis (EDRDA) for structured road environments~\cite{yenikaya2013keeping}. The lane tracking algorithm then monitors and updates the position of detected lanes in real-time using a predefined motion model. This model predicts the future position of the vehicle based on parameters such as current speed, direction, and steering inputs~\cite{kumar2015review}. In the second step, the planning layer calculates an optimal future trajectory~\cite{motionplanningmodule2023}. Lastly, the control layer executes the trajectory produced by the planner as actuation commands in the form of steering angle changes. The range of these changes is limited by the physical constraints of mechanical control units to ensure driving stability and safety~\cite{becker2018functional}.

\section{Related Work}
\label{sec:3-Related_Work}
Several studies focus on attacking ALC systems that affect AV navigation. Boloor \textit{et al.}~\cite{boloor2019simple} showed an end-to-end attack using simple physical manipulations, such as painting black lines on roads, to alter AV paths. The Black Strip Attack by Boloor \textit{et al.} is a direct legal violation, as it involves painting a dark patch on the street~\cite{shouselaw2024}. However, the attack is practical and stealthy, requiring only basic materials to visually alter the road while mimicking road imperfections, such as tire skid marks or dirt. Sato \textit{et al.}~\cite{sato2021dirty} proposed the DRP-attack, an adversarial perturbation on a road patch that drives an AV off-road and requires printing a long patch tailored for each AV and posing as street workers to apply it. This is difficult on busy roads and violates the law because attackers disguise themselves to place the patch. Such actions, including painting or altering road surfaces without authorization, are illegal~\cite{usdot2023}.

Jing \textit{et al.}~\cite{jing2021too} proposed a two-stage attack that first solves an optimization problem for perturbation placement in the digital domain and then maps the results to physical-world markings, misleading Tesla Autopilot's lane detection. However, this attack assumes the adversary can reverse engineer and instrument Tesla Autopilot to extract internal camera and lane images and run the optimization per scenario, which limits its practicality and it involves illegal alteration of public road surfaces~\cite{usdot2023}. Cao \textit{et al.}~\cite{caoremote} conducted an attack on ALC by placing an adversarial patch on the back of a vehicle ahead of the victim car. The RAP-ALC attack by Cao \textit{et al.} uses adversarial patches placed behind a vehicle. Although practical, as it does not require presence in the road, the large visible patch reduces stealth. The attacker can execute it from their own vehicle, placing it in a \textit{legal gray area}, since it does not involve directly altering public infrastructure, but it is still visible to the victim, reducing effectiveness. \autoref{tab:attacks_comparison} compares attacks targeting the ALC module, highlighting the operational criteria of each.

\section{Threat Model and Attack Objectives}
\label{sec:4-Attack_Formulation}
\subsection{Adversary Assumptions and Scope}

\noindent \textbf{Attacker Goals.} The attacker's primary objective is to deceive the perception system of a victim SAE Level 1--2 vehicle, causing misdetection of \NSes as genuine lane markings and leading to deviation from the intended path. Such deviations pose severe safety hazards, including:
\begin{enumerate}[noitemsep, nolistsep, label=\arabic{enumi}.,ref=Step \arabic{enumi}, leftmargin=*]
\item \textbf{Lane departure into oncoming traffic}: Increases collision risk with vehicles or bicycles, potentially causing accidents and vehicle damage.
\item \textbf{Road departure into infrastructure}: Results in the vehicle running over curbs and colliding with roadside objects such as bus stations or pedestrians, posing serious public safety threats.
\end{enumerate}

\noindent \textbf{Attacker Capabilities.} To execute the \NSAttack, the attacker uses an object, such as a canopy, placed on their own property, as illustrated in \autoref{fig:teaser_picture}. This canopy casts a shadow onto the street while forming two bright areas within it, created by sunlight passing through two rectangular holes. These bright areas resemble genuine line markings. Because shadows are natural and ubiquitous environmental phenomena, the attack is subtle and difficult to detect.

A key advantage of this method is that the attacker does not need to manipulate public infrastructure or directly violate traffic laws. Furthermore, the attacker does not require specific knowledge of the victim's AV perception algorithm. Instead, the attack exploits vulnerabilities shared across LD architectures, as demonstrated by our successful attacks against three diverse models: CLRerNet, HybridNets, and TwinLiteNet. As highlighted in~\autoref{tab:attacks_comparison}, the \NSAttack meets operational criteria while remaining challenging to detect.

\begin{figure}[t]
    \centering
    \includegraphics[width=0.7\columnwidth]{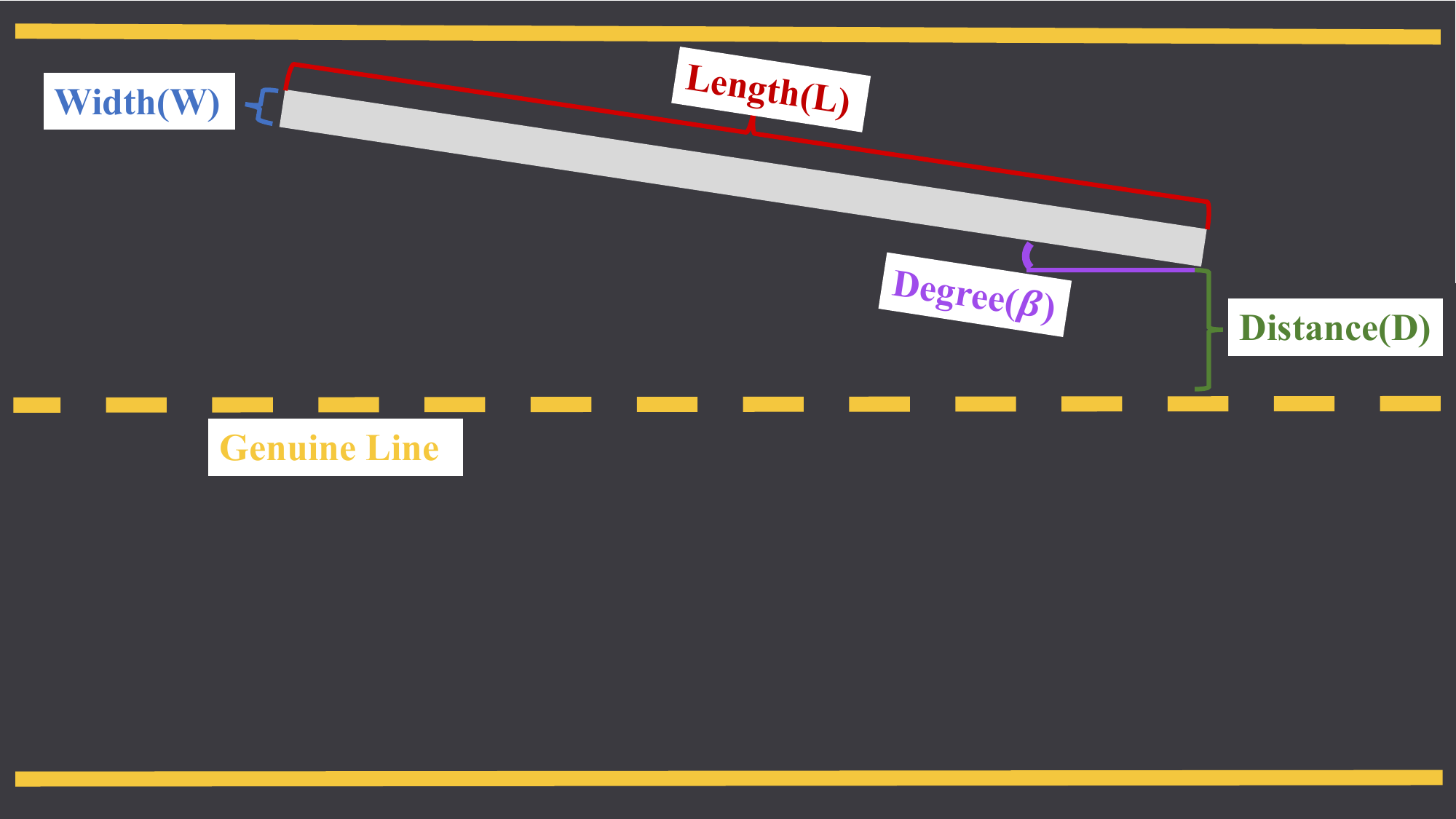}
    \caption{Key parameters of a \NS: width (\(W\)), length (\(L\)), lateral distance from line (\(D\)), and angle (\(\beta\)).}
    \label{fig:parameters}
\end{figure}

\begin{figure*}[!t]
\centering

\subfloat[CLRerNet]{%
    \resizebox{0.24\textwidth}{0.13\textheight}{%
    \begin{tikzpicture}
      \foreach \x in {0,3,6,9} \draw (\x,0) -- (\x,5);
      \foreach \y in {0,1,2,3,4,5} {
        \draw[gray!30] (0,\y) -- (9,\y);

        \pgfmathsetmacro{\WW}{\y*72}
        \draw (0.1,\y) -- (-0.1,\y)
          node[left]{\textbf{\pgfmathprintnumber[precision=0]{\WW}}};

        \pgfmathsetmacro{\LL}{\y*8}
        \draw (2.9,\y) -- (3.1,\y)
          node[left] at (2.85,\y){\textbf{\pgfmathprintnumber[precision=0]{\LL}}};

        \pgfmathsetmacro{\AA}{\y*18}
        \draw (5.9,\y) -- (6.1,\y)
          node[left] at (5.85,\y){\textbf{\pgfmathprintnumber[precision=0]{\AA}}};

        \pgfmathsetmacro{\DD}{\y*70}
        \draw (8.9,\y) -- (9.1,\y)
          node[left] at (8.85,\y){\textbf{\pgfmathprintnumber[precision=0]{\DD}}};
      }
      \node[font=\large\bfseries] at (0,5.7){Width (cm)};
      \node[font=\large\bfseries] at (3,5.7){Length (m)};
      \node[font=\large\bfseries] at (6,5.7){Angle ($^\circ$)};
      \node[font=\large\bfseries] at (9,5.7){Distance (cm)};
      \input{Plots/clrernet_lanes.tex}
    \end{tikzpicture}%
  }\label{subfig:CLRerNet}}
\hfill
\subfloat[HybridNet]{%
    \resizebox{0.24\textwidth}{0.13\textheight}{%
    \begin{tikzpicture}
      \foreach \x in {0,3,6,9} \draw (\x,0) -- (\x,5);
      \foreach \y in {0,1,2,3,4,5} {
        \draw[gray!30] (0,\y) -- (9,\y);

        \pgfmathsetmacro{\WW}{\y*72}
        \draw (0.1,\y) -- (-0.1,\y)
          node[left]{\textbf{\pgfmathprintnumber[precision=0]{\WW}}};

        \pgfmathsetmacro{\LL}{\y*8}
        \draw (2.9,\y) -- (3.1,\y)
          node[left] at (2.85,\y){\textbf{\pgfmathprintnumber[precision=0]{\LL}}};

        \pgfmathsetmacro{\AA}{\y*18}
        \draw (5.9,\y) -- (6.1,\y)
          node[left] at (5.85,\y){\textbf{\pgfmathprintnumber[precision=0]{\AA}}};

        \pgfmathsetmacro{\DD}{\y*70}
        \draw (8.9,\y) -- (9.1,\y)
          node[left] at (8.85,\y){\textbf{\pgfmathprintnumber[precision=0]{\DD}}};
      }
      \node[font=\large\bfseries] at (0,5.7){Width (cm)};
      \node[font=\large\bfseries] at (3,5.7){Length (m)};
      \node[font=\large\bfseries] at (6,5.7){Angle ($^\circ$)};
      \node[font=\large\bfseries] at (9,5.7){Distance (cm)};
      \input{Plots/hybridnet_lanes.tex}
    \end{tikzpicture}%
  }\label{subfig:hybridnet}}
\hfill
\subfloat[TwinLiteNet]{%
    \resizebox{0.24\textwidth}{0.13\textheight}{%
    \begin{tikzpicture}
      \foreach \x in {0,3,6,9} \draw (\x,0) -- (\x,5);
      \foreach \y in {0,1,2,3,4,5} {
        \draw[gray!30] (0,\y) -- (9,\y);

        \pgfmathsetmacro{\WW}{\y*72}
        \draw (0.1,\y) -- (-0.1,\y)
          node[left]{\textbf{\pgfmathprintnumber[precision=0]{\WW}}};

        \pgfmathsetmacro{\LL}{\y*8}
        \draw (2.9,\y) -- (3.1,\y)
          node[left] at (2.85,\y){\textbf{\pgfmathprintnumber[precision=0]{\LL}}};

        \pgfmathsetmacro{\AA}{\y*18}
        \draw (5.9,\y) -- (6.1,\y)
          node[left] at (5.85,\y){\textbf{\pgfmathprintnumber[precision=0]{\AA}}};

        \pgfmathsetmacro{\DD}{\y*70}
        \draw (8.9,\y) -- (9.1,\y)
          node[left] at (8.85,\y){\textbf{\pgfmathprintnumber[precision=0]{\DD}}};
      }
      \node[font=\large\bfseries] at (0,5.7){Width (cm)};
      \node[font=\large\bfseries] at (3,5.7){Length (m)};
      \node[font=\large\bfseries] at (6,5.7){Angle ($^\circ$)};
      \node[font=\large\bfseries] at (9,5.7){Distance (cm)};
      \input{Plots/twinlitenet_lanes.tex}
    \end{tikzpicture}%
  }\label{subfig:twinlitenet}}
\hfill
\subfloat[Consensus Region\label{fig:consensus}]{%
  \resizebox{0.24\textwidth}{0.13\textheight}{%
    \begin{tikzpicture}
      \foreach \x in {0,3,6,9} \draw (\x,0) -- (\x,5);
      \foreach \y in {0,1,2,3,4,5} {
        \draw[gray!30] (0,\y) -- (9,\y);

        \pgfmathsetmacro{\WW}{\y*72}
        \draw (0.1,\y) -- (-0.1,\y)
          node[left]{\textbf{\pgfmathprintnumber[precision=0]{\WW}}};

        \pgfmathsetmacro{\LL}{\y*8}
        \draw (2.9,\y) -- (3.1,\y)
          node[left] at (2.85,\y){\textbf{\pgfmathprintnumber[precision=0]{\LL}}};

        \pgfmathsetmacro{\AA}{\y*18}
        \draw (5.9,\y) -- (6.1,\y)
          node[left] at (5.85,\y){\textbf{\pgfmathprintnumber[precision=0]{\AA}}};

        \pgfmathsetmacro{\DD}{\y*70}
        \draw (8.9,\y) -- (9.1,\y)
          node[left] at (8.85,\y){\textbf{\pgfmathprintnumber[precision=0]{\DD}}};
      }
      \node[font=\large\bfseries] at (0,5.7){Width (cm)};
      \node[font=\large\bfseries] at (3,5.7){Length (m)};
      \node[font=\large\bfseries] at (6,5.7){Angle ($^\circ$)};
      \node[font=\large\bfseries] at (9,5.7){Distance (cm)};
      \input{Plots/intersection_lanes.tex}
    \end{tikzpicture}%
  }}
\caption{Successful \NS attack ranges for each model, and the consensus region where all three misdetect the \NS.}
\label{fig:syseval_pc}
\end{figure*}

\noindent \textbf{Constraints.} The attack is most effective when the property faces east or west and the road runs north--south, a common configuration in grid-based U.S. city layouts~\cite{boeing2018}. This geometry allows sufficiently long shadows during daylight hours to affect vehicles traveling at legal speed limits. According to the 2017 American Housing Survey, 73\% of U.S. households are located in suburban or rural areas~\cite{huduser2017}, where properties often provide enough space for installing a canopy or fence without drawing attention. For example, at 34\textdegree N latitude, a 10\,m-long object mounted at 10\,m height can cast shadows ranging from roughly 20\,m near midday to over 60\,m when the Sun is low, depending on seasonal and daily sun angles; detailed calculations are provided in Section~\ref{sec:5_2_Shadow_Calculation}. Since the object remains on private property and does not physically alter public roads, the attack operates in a legal gray zone: it creates hazardous driving conditions indirectly, without overtly violating traffic regulations.

\subsection{Objectives and Legal Considerations}

We define three key considerations that guided the \NSAttack design to ensure feasibility, legality, and stealth.

\noindent \textbf{C1. LD algorithms incorporate shadow-removal pre-processing that introduces exploitable intensity bias.}
Shadows caused by objects blocking sunlight can fragment lane markings or alter their apparent brightness, causing misdetection~\cite{zhao2022lane}. To counter this, LD pipelines apply shadow-removal techniques, including color-space transformations and histogram equalization~\cite{liu2021condlanenet}. However, these methods can introduce an intensity bias: bright regions within shadowed areas may be treated as non-shadow regions or lane-like structures. The \NSAttack directly exploits this weakness by creating bright lane-like regions inside a shadowed area.

\noindent \textbf{C2. The attack must be deployable from private property, without on-road presence or legally prohibited modifications.}
Physical attacks such as road painting~\cite{boloor2019simple}, adversarial patches~\cite{sato2021dirty}, or impersonating road workers require direct roadway access and may violate traffic laws~\cite{uslegal2023criminalimpersonation}. They also require prolonged on-road presence, increasing detection risk. In contrast, the \NSAttack requires only an off-road object, such as a canopy, placed on the attacker's own property. This avoids direct public-road modification and reduces the need for risky on-road operations.

\noindent \textbf{C3. If a human is in the loop, obvious attacks will result in manual takeover and mitigation.}
In SAE Levels 1--2, where human drivers must remain alert and prepared to take control, visible anomalies can trigger immediate manual intervention, mitigating the attack's impact. Therefore, for the attack to succeed, the induced pattern must remain visually inconspicuous to human drivers.

\section{Negative Shadow Attack Design}
\label{sec:5-Negative_Shadow_Attack_Design}
\subsection{Attack Overview and Physical Basis}
\label{sec:5_1_Negative_Shadow}

To address \textbf{C1}, we developed the \NSAttack, which leverages the concept of \emph{negative shadows}: deceptive bright patterns on the road that resemble genuine lane markings. These patterns are created by sunlight passing through openings in an opaque light barrier, such as a canopy, positioned on private property adjacent to the road. The openings are designed to project bright lane-like regions inside a larger shadowed area. As a result, the attack can induce lane-misdetection without physically altering the road surface or placing any object in the lane.

The \NSAttack exploits \emph{intensity bias} in LD algorithms, i.e., the tendency of shadow-removal or illumination-normalization pre-processing to rely heavily on pixel brightness when distinguishing shadowed from non-shadowed regions~\cite{zhu2021mitigating}. In this setting, bright regions inside a shadowed area may be treated as non-shadow regions and preserved by the pre-processing pipeline. Because the resulting \NSes have lane-like shape, contrast, and orientation, LD algorithms may misclassify them as genuine lane markings. This vulnerability manifests differently across algorithm types:

\begin{enumerate}[noitemsep, nolistsep, label=(\roman*)]
\item \textbf{Feature-based LD algorithms}, such as those using gradient-based methods or histogram-based thresholding, are vulnerable because \NSes mimic the visual attributes of genuine lane markings~\cite{wang2010model}. Their sharp edges and strong local contrast can trigger responses to abrupt pixel-intensity changes, leading to false line classifications in binary segmentation outputs~\cite{kim2003high, wang2010model}. This increases the likelihood that the perception system treats the \NSes as valid lane boundaries.

\item \textbf{Model-based LD algorithms}, which use polynomial regression or spline models to estimate lane geometry, are also susceptible because \NSes can introduce spurious lane-like evidence into the fitting process~\cite{davies2004machine}. When \NSes are positioned close to actual lane markings or aligned with the expected lane direction, they can distort the estimated lane model, producing erroneous trajectories~\cite{piccioli1996robust}.
\end{enumerate}

Because the projected patterns resemble naturally occurring illumination artifacts rather than painted or attached road markings, the attack also supports \textbf{C3}: it reduces the likelihood that a human driver notices the manipulation and performs a manual takeover.

\subsection{Geometric Design of Shadows}
\label{sec:5_2_Shadow_Calculation}

\noindent \textbf{Sun--Canopy Shadow Geometry.}
The ground-plane shadow length $SL$ cast by a canopy of length $L_{\text{canopy}}$
mounted at height $h$ on a building facade is governed by the Sun's altitude $ALT$
and azimuth $AZ$:
\begin{equation}
SL = \frac{h}{\tan(ALT)} \;+\; L_{\text{canopy}} \cos(AZ - \theta).
\end{equation}
where $\theta$ is the canopy's orientation relative to the road. When aligned
with the Sun ($AZ \approx \theta$), the full canopy length contributes to $SL$.
For example, a 5\,m canopy on a 10\,m building yields $SL \approx 15$\,m at
solar noon ($ALT=45^\circ$) and $SL \approx 62$\,m at low sun ($ALT=10^\circ$),
demonstrating that short canopies can produce shadows spanning tens of meters. The detailed solar-position calculations and five-year solar-feasibility analysis are provided in Appendix~\ref{sec:11-Appendix_A}.

\noindent \textbf{Dynamic Deployment and Legal Basis.}
To address \textbf{C2}, the \NSAttack is implemented using physical components that remain off the lane surface. An opaque canopy or panel mounted on a movable holder can be positioned on private property to cast \NSes onto the roadway without requiring the attacker to enter the lane, paint the road, attach patches to the pavement, or impersonate road workers. Given the Sun's position and the road geometry, the attacker only needs to select the holder location and orientation so that the projected bright regions mimic lane markings.

This deployment mechanism avoids direct manipulation of public infrastructure and reduces the need for prolonged on-road operations. Since the object remains on the attacker's property, the attack operates in a legal gray zone: it indirectly creates hazardous driving conditions while avoiding overt traffic-law violations associated with modifying the road surface. The setup is also consistent with typical property setback rules for structures near property lines~\cite{rockethomes2023propertyline}.

\subsection{Parameterization and Test Configuration}
\label{sec:parameters_scenarios}

\begin{algorithm}[t]
\caption{Search-Based \NSAttack via Genetic Algorithm}
\label{alg:search_based_ns}
\DontPrintSemicolon
\scriptsize
\SetNlSty{}{}{: }

\textbf{Input:} $N$, $G$, $I_{\mathrm{BEV}}$, $H_{\mathrm{BEV}\rightarrow\mathrm{DRV}}$, $\Delta L^*$, $\tau$\;
Initialize $N$ candidates $(W,L,D,\beta)$ uniformly within the predefined bounds\;

\For{$g=1$ \KwTo $G$}{
    Construct each valid $\mathcal{P}_{\mathrm{BEV}}$ and add $\Delta L^*$ inside the \NS region\;
    Warp the modified image and $\mathcal{P}_{\mathrm{BEV}}$ to the driver view using $H_{\mathrm{BEV}\rightarrow\mathrm{DRV}}$\;
    Evaluate TwinLiteNet, HybridNets, and CLRerNet and compute $\gamma_k$ and $\delta_k$\;
    Compute $f_i=0.5\,\mathrm{avg}_k(\gamma_k)+0.5\,\mathrm{avg}_k(\delta_k)$\;
    Select the top $N/2$ candidates and apply crossover and mutation\;
}
\textbf{Return:} $(W^*,L^*,D^*,\beta^*)$ with maximum fitness\;
\end{algorithm}

We define four parameters for constructing a \NS. \textbf{Width} \(W\) is the width of the bright stripe. \textbf{Length} \(L\) determines how far the \NS extends longitudinally along the road. \textbf{Lateral distance} \(D\) is the distance between the lane marking and the \NS. \textbf{Angle} \(\beta\) is the orientation of the \NS relative to the lane marking. These parameters, illustrated in \autoref{fig:parameters}, are used to design and evaluate the impact of the \NSAttack through three safety-critical scenarios.

\noindent\textbf{Scenario 1 (Left-turn Maneuver Attack).}
The victim vehicle approaches an intersection intending to turn left, as shown in \autoref{fig:first_scenario}. By aligning the \NSes to appear like a continuation of the lane rather than a turn, the LD system may fail to initiate the left-turn maneuver. This misdetection can cause the vehicle to continue straight, potentially leading to collisions with barriers or off-road hazards.

\noindent\textbf{Scenario 2 (Head-on Collision Attack).}
On a local two-way road, as shown in \autoref{fig:second_scenario}, the adversary positions \NSes so that the victim vehicle drifts into the oncoming lane. When the \NSes are mistaken for lane boundaries, the ALC system may steer the vehicle toward an approaching vehicle, increasing the risk of a head-on collision.

\noindent\textbf{Scenario 3 (Off-road Deviation Attack).}
As shown in \autoref{fig:third_scenario}, the attacker creates \NSes along the right shoulder of the road. If these patterns are misdetected as lane markings, the vehicle may shift its perceived drivable space toward the shoulder and swerve off-road. This can result in dangerous run-off scenarios, including collisions with roadside infrastructure such as bus stops or pedestrians.

Overall, these three scenarios highlight the safety risks introduced by the \NSAttack. Subsequent sections evaluate how misdetection of \NSes affects LD outputs and leads to unsafe vehicle maneuvers in simulation, miniature testbed experiments, and real-world deployment.

\section{Evaluation of Lane Detection Misdetection}
\label{sec:6-Misdetection_Evaluation}
\noindent \textbf{\NSAttack Evaluation.} \label{sec:ns_attack_evaluation} This section evaluates how three LD algorithms --- \textit{CLRerNet}~\cite{honda2023clrernet}, \textit{TwinLiteNet}~\cite{che2023twinlitenet}, and \textit{HybridNets}~\cite{vu2022hybridnets} --- respond to \NSAttack characterized by their width $(W)$, length $(L)$, lateral distance $(D)$, and angle $(\beta)$. These models were selected based on their high reported accuracy on public leaderboards~\cite{paperswithcode2024}, with \textit{TwinLiteNet} representing a Feature-Based approach, and \textit{HybridNets} and \textit{CLRerNet} combining Feature-Based and Model-Based components. All experiments were conducted on scenes with line markings, although the \NSAttack may be more effective in their absence. Experiments were conducted using Python 3.8 on Ubuntu 22.04 with 64 GB RAM, an NVIDIA H100 80 GB GPU, and an Intel i9-13900KF CPU.

\begin{figure}[t]
    \centering
    %\vspace{-3mm} % optional
    \begin{tikzpicture}
        \begin{axis}[
            width=0.6\columnwidth,
            height=3.5cm,
            xlabel={Negative Shadow Length (m)},
            ylabel={Success rate (\%)},
            ymin=0, ymax=105,
            xtick={25,30,35},
            ytick={0,20,...,100},
            xticklabel style={font=\small},
            yticklabel style={font=\small},
            label style={font=\small},
            grid=major,
            grid style={dashed,gray!30},
            legend style={
                at={(1.32,0.1)},
                anchor=south,
                font=\scriptsize,
            },
            legend columns=1,
            legend cell align=left,
            mark options={scale=1.3},
        ]

        % Scenario 1
        \addplot+[mark=square*, thick, blue] coordinates {
            (25,100) (30,100) (35,100)
        };

        % Scenario 2
        \addplot+[mark=triangle*, thick, red] coordinates {
            (25,80) (30,100) (35,100)
        };

        % Scenario 3
        \addplot+[mark=o, thick, green!60!black] coordinates {
            (25,40) (30,60) (35,80)
        };

        % Average
        \addplot+[mark=*, thick, black, dashed] coordinates {
            (25,73.333) (30,86.667) (35,93.333)
        };

        \legend{Scenario 1, Scenario 2, Scenario 3, Average}

        \end{axis}
    \end{tikzpicture}
    \caption{Success rate of the \NegativeShadow attack for different NS lengths across the three scenarios.}
    \label{fig:success_rate}
\end{figure}

The \NS generation algorithm~\ref{alg:search_based_ns} begins with a BEV image and generates candidate \NegativeShadow parameterized by \(W\), \(L\), \(D\), and \(\beta\),  with geometric parameters constrained to keep \NS within one lane. These values are sampled uniformly within predefined bounds to ensure each polygon \(\mathcal{P}_{\text{BEV}}\) is convex. For each candidate, a binary mask is applied to the BEV image, which is first converted from \(RGB\) to \(LAB\) color space---allowing targeted adjustment of the luminance channel \(L^*(x,y)\) while preserving color integrity. Brightening is achieved by increasing \(L^*(x,y)\) by a fixed offset \(\Delta_L\) wherever \((x,y) \in \mathcal{P}_{\text{BEV}}\), which was empirically determined through preliminary experiments that setting $\Delta_L = 25$ best matches the average $L^*$ difference observed between sunlit and shaded regions, thus producing realistic brightness shifts. This photometric modification is represented by the function:
\begin{equation}
F_{\text{NS}}\bigl(I_{\text{BEV}}, W, D, L, \beta\bigr)
\end{equation}
which returns the modified BEV image and polygon coordinates. The image is then warped to driver view via a fixed homography \(H_{\text{BEV} \rightarrow \text{DRV}}\), and passed through three LD models. For each model \(\mathcal{LD}_i\), we compute an overlap ratio \(\gamma_i\) between the predicted line mask \(M_{\mathcal{LD}_i}\) and the projected \NS polygon \(\mathcal{P}_{\text{DRV}}\):
\begin{equation}
  \gamma_i = \frac{|\mathcal{P}_{\text{DRV}} \cap M_{\mathcal{LD}_i}|}{|\mathcal{P}_{\text{DRV}}|}.
\end{equation}

Following the evaluation of each \NS candidate, its projected polygon \(\mathcal{P}_{\text{DRV}}\) is used to compute geometric attributes including length \(L_i\), width \(W_i\), lateral distance \(D_i\), and incidence angle \(\beta_i\). These quantities are measured in meters or degrees by converting from pixel coordinates using a fixed meters-per-pixel ratio \(M_{\text{px}}\). %For instance,~\autoref{fig:width_distance_combined} illustrates how all three detectors simultaneously misinterpret a 40m-long, width-matched NS stripe positioned just 0.1m from the genuine line marking.
These overlaps are combined with geometric features into a composite fitness function:
\begin{equation}
f_i = 0.5 \cdot \operatorname{avg}(\gamma_1, \gamma_2, \gamma_3)
    + 0.5 \cdot \operatorname{avg}(\delta_1, \delta_2, \delta_3),
\end{equation}
where \(\gamma_k\) denotes overlap and \(\delta_k\) indicates misclassification. The resulting score evaluates each \NS candidate based on both detection strength and detection consistency---(1) average overlap with LD model outputs and (2) the proportion of LD models that misclassify the \NS as a line---encouraging configurations that consistently induce errors. This score guides the evolutionary search across generations. 

The overlap term \(\gamma_k\) quantifies how strongly the \NS confuses each LD, while the binary detection term \(\delta_k\) captures whether the \NS is misclassified as a line by each model. This formulation encourages \NegativeShadows that produce strong spatial overlap with predicted lines, increasing the chance of misclassification. Conceptually, the entire loop is governed by two abstract functions. The LD operator

%%%%%%
\begin{figure}[t]
  \centering
  % Row 1
  \begin{subfigure}[b]{0.47\linewidth}
    \centering
    \fbox{\includegraphics[width=\linewidth]{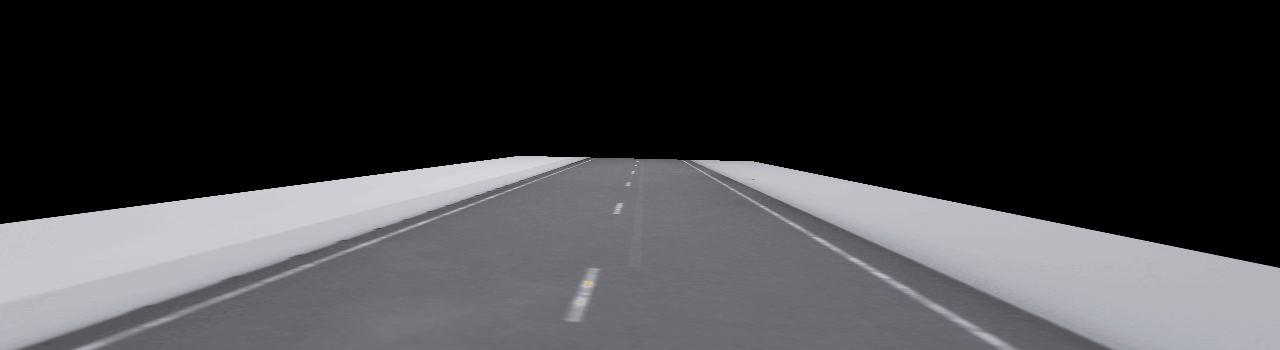}}
    \caption{$\Delta L^* = 5$}
  \end{subfigure}
  \hfill
  \begin{subfigure}[b]{0.47\linewidth}
    \centering
    \fbox{\includegraphics[width=\linewidth]{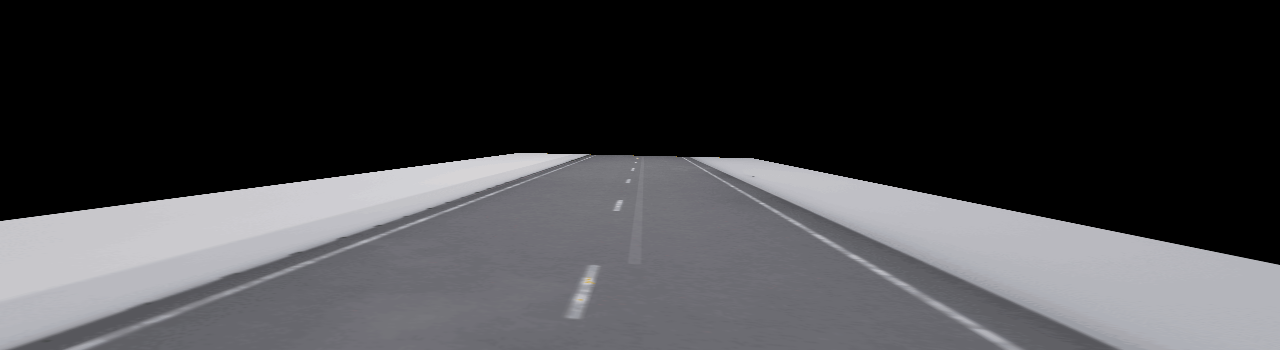}}
    \caption{$\Delta L^* = 10$}
  \end{subfigure}

  \par\medskip

  % Row 2
  \begin{subfigure}[b]{0.47\linewidth}
    \centering
    \fbox{\includegraphics[width=\linewidth]{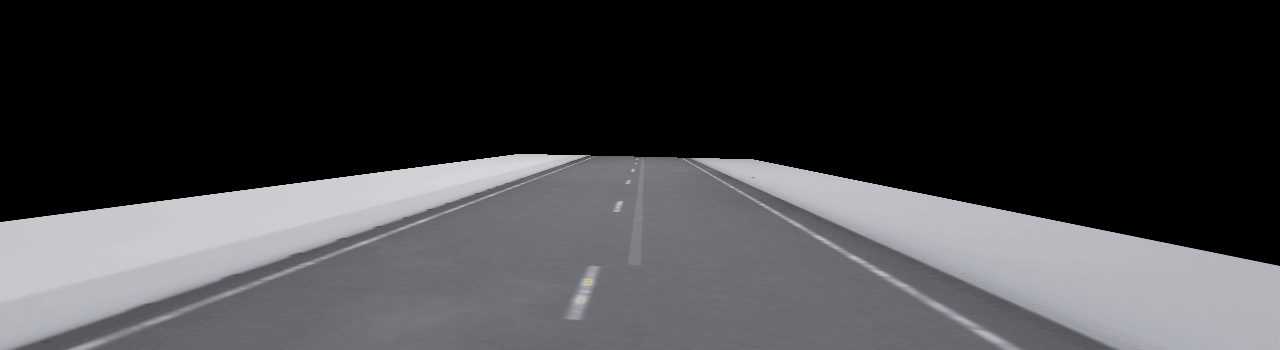}}
    \caption{$\Delta L^* = 15$}
  \end{subfigure}
  \hfill
  \begin{subfigure}[b]{0.47\linewidth}
    \centering
    \fbox{\includegraphics[width=\linewidth]{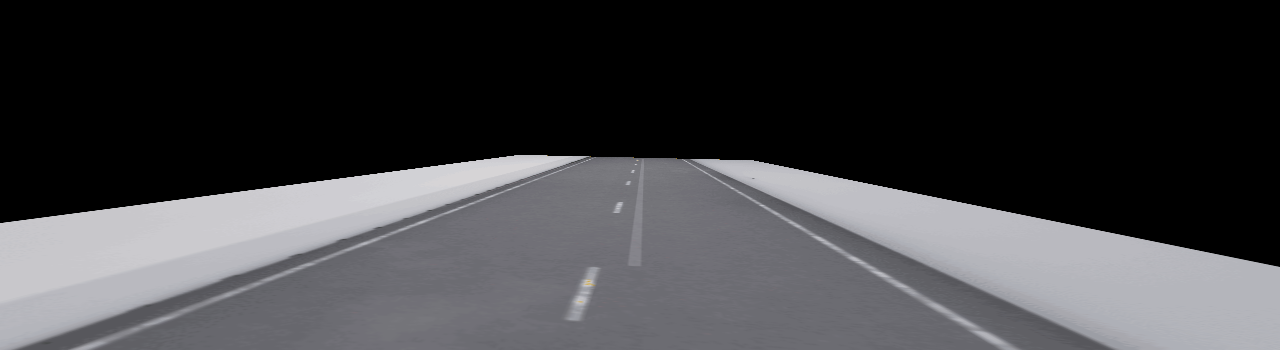}}
    \caption{$\Delta L^* = 20$}
  \end{subfigure}

  \par\medskip

  % Row 3
  \begin{subfigure}[b]{0.47\linewidth}
    \centering
    \fbox{\includegraphics[width=\linewidth]{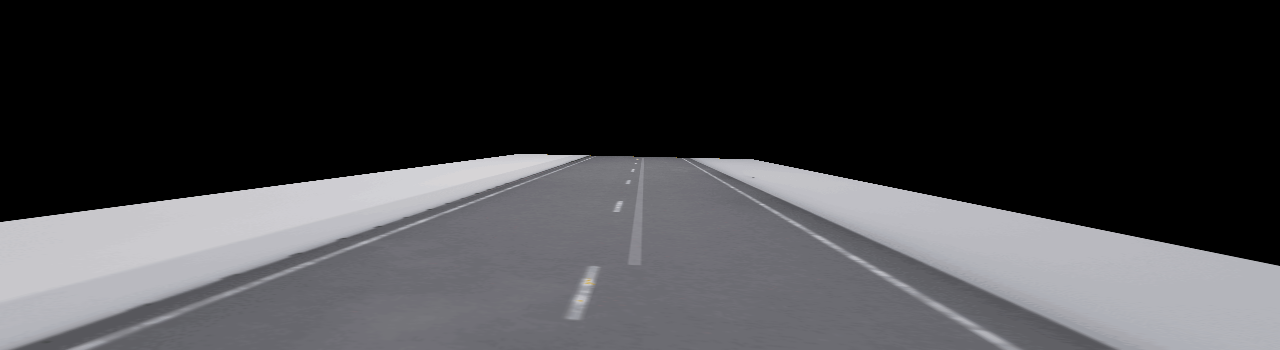}}
    \caption{$\Delta L^* = 25$}
  \end{subfigure}
  \hfill
  \begin{subfigure}[b]{0.47\linewidth}
    \centering
    \fbox{\includegraphics[width=\linewidth]{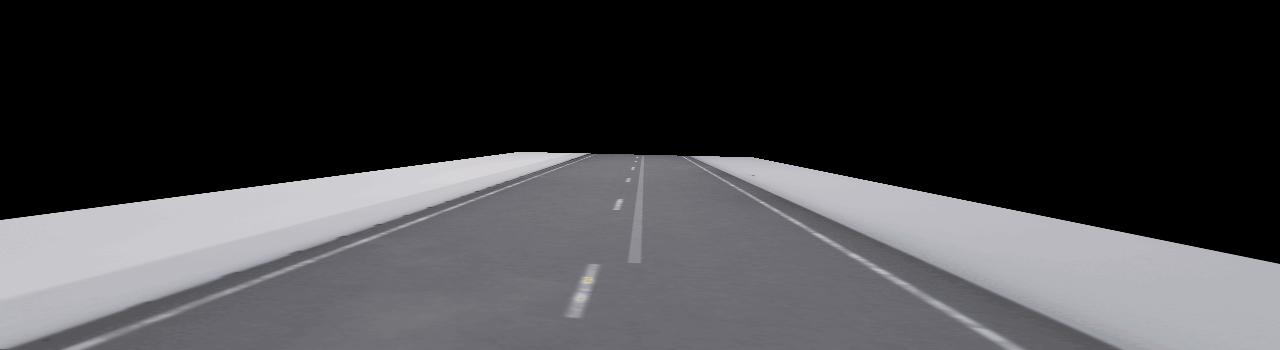}}
    \caption{$\Delta L^* = 30$}
  \end{subfigure}
  \caption{Representative calibrated \NS renderings for six illumination settings
  $\Delta L^* \in \{5,10,15,20,25,30\}$, obtained by matching \NS--road contrast
  to CARLA daytime presets.}
  \label{fig:illumination_carla}
\end{figure}

%%%%%%
\begin{figure*}[t]
  \centering
  \begin{subfigure}[t]{0.25\textwidth}
    \centering
    \resizebox{\linewidth}{!}{% ---------- LENGTH (L) ABLATION ----------
\begin{tikzpicture}[scale=2, rotate=180]

% --- Basic radii ---
\def\rInner{0.2}      % inner hole radius
\def\ringWidth{0.3}   % thickness of each ring
\def\nRings{4}        % 4 rings -> 5 circles including inner

% Outer radius of the last circle
\pgfmathsetmacro{\rOuter}{\rInner + \nRings*\ringWidth}

% ==========================================================
% Ring-aware cell fill:
% ring 1..4 = L=10,20,30,40 (inner -> outer)
% ==========================================================
\newcommand{\FillCellRing}[4]{%
  % #1 = start hour, #2 = end hour, #3 = ring index 1..4, #4 = color
  \pgfmathsetmacro{\startAng}{90 - 15*#1}
  \pgfmathsetmacro{\endAng}{90 - 15*#2}
  \pgfmathsetmacro{\rA}{\rInner + (#3-1)*\ringWidth}
  \pgfmathsetmacro{\rB}{\rInner + (#3)*\ringWidth}
  \fill[#4, opacity=0.75]
    (\startAng:\rA) -- (\startAng:\rB)
      arc[start angle=\startAng, end angle=\endAng, radius=\rB]
    -- (\endAng:\rA)
      arc[start angle=\endAng, end angle=\startAng, radius=\rA]
    -- cycle;
}

\newcommand{\FillCellAllL}[6]{%
  % #1,#2 = start/end hour
  % #3..#6 = colors for L=10, L=20, L=30, L=40 (inner->outer)
  \FillCellRing{#1}{#2}{1}{#3}%
  \FillCellRing{#1}{#2}{2}{#4}%
  \FillCellRing{#1}{#2}{3}{#5}%
  \FillCellRing{#1}{#2}{4}{#6}%
}

% ==========================================================
% Fills from Length-lux-* table
% ==========================================================

% ΔL=5:  L10=RED, L20=RED, L30=YELLOW, L40=YELLOW
\FillCellAllL{6}{7}{red!70}{red!70}{yellow!80}{yellow!80}
\FillCellAllL{17}{18}{red!70}{red!70}{yellow!80}{yellow!80}

% ΔL=10: L10=RED, L20=YELLOW, L30=GREEN, L40=GREEN
\FillCellAllL{7}{8}{red!70}{yellow!80}{green!60}{green!60}
\FillCellAllL{16}{17}{red!70}{yellow!80}{green!60}{green!60}

% ΔL=15..35: all GREEN
\FillCellAllL{8}{9}{green!60}{green!60}{green!60}{green!60}
\FillCellAllL{9}{10}{green!60}{green!60}{green!60}{green!60}
\FillCellAllL{10}{11}{green!60}{green!60}{green!60}{green!60}
\FillCellAllL{11}{12}{green!60}{green!60}{green!60}{green!60}
\FillCellAllL{12}{13}{green!60}{green!60}{green!60}{green!60}
\FillCellAllL{13}{14}{green!60}{green!60}{green!60}{green!60}
\FillCellAllL{14}{15}{green!60}{green!60}{green!60}{green!60}
\FillCellAllL{15}{16}{green!60}{green!60}{green!60}{green!60}

% --- 5 nested half circles (semi-doughnut) ---
\foreach \k in {0,...,\nRings} {
  \pgfmathsetmacro{\rr}{\rInner + \k*\ringWidth}
  \draw[thick] (0:\rr) arc[start angle=0, end angle=-180, radius=\rr];
}
% close the annulus edges on the diameter
\draw[thick] (0:\rInner) -- (0:\rOuter);
\draw[thick] (-180:\rInner) -- (-180:\rOuter);

% --- L labels on the 12:00 hour line (inner->outer: 10,20,30,40) ---
\pgfmathsetmacro{\angTwelve}{90 - 15*12}
\foreach[count=\i from 1] \ell in {10,20,30,40}{%
  \pgfmathsetmacro{\rMid}{\rInner + (\i-0.5)*\ringWidth}
  \node[font=\footnotesize] at (\angTwelve:\rMid)
    {\textbf{$L{=}\ell\,\mathrm{m}$}};
}

% --- Hour rays + labels for the half (6..18) ---
\foreach \h in {6,...,18} {
  \pgfmathsetmacro{\ang}{90 - 15*\h}
  \draw[gray!60] (\ang:\rInner) -- (\ang:{\rOuter+0.10});
  \node[font=\tiny] at (\ang:{\rOuter+0.18}) {\h};
}

% --- Radius for the dashed double-headed arrows ---
\pgfmathsetmacro{\rArrow}{\rOuter + 0.35}

% ==========================================================
% ΔL arc labels — shortened to number only to avoid overlap
% ==========================================================
\newcommand{\DeltaLArc}[3]{%
  % #1 = ΔL value, #2 = start hour, #3 = end hour
  \pgfmathsetmacro{\startAng}{90 - 15*#2}%
  \pgfmathsetmacro{\endAng}{90 - 15*#3}%
  \draw[<->, dashed, thick]
    (\startAng:\rArrow)
      arc[start angle=\startAng, end angle=\endAng, radius=\rArrow]
      node[midway, sloped, allow upside down=false, above, yshift=0.5pt,
           font=\scriptsize\bfseries, fill=white, inner sep=0.5pt] {$#1$};%
}

% --- ΔL arcs ---
\DeltaLArc{5}{6}{7}
\DeltaLArc{10}{7}{8}
\DeltaLArc{15}{8}{9}
\DeltaLArc{20}{9}{10}
\DeltaLArc{25}{10}{11}
\DeltaLArc{30}{11}{12}
\DeltaLArc{30}{12}{13}
\DeltaLArc{25}{13}{14}
\DeltaLArc{20}{14}{15}
\DeltaLArc{15}{15}{16}
\DeltaLArc{10}{16}{17}
\DeltaLArc{5}{17}{18}

\end{tikzpicture}}
    \caption{Length ($L$)}
    \label{fig:ablation_length}
  \end{subfigure}\hfill
  \begin{subfigure}[t]{0.25\textwidth}
    \centering
    \resizebox{\linewidth}{!}{% ---------- WIDTH (W) ABLATION ----------
\begin{tikzpicture}[scale=2, rotate=180]

% --- Basic radii ---
\def\rInner{0.2}      % inner hole radius
\def\ringWidth{0.3}   % thickness of each ring
\def\nRings{4}        % 4 rings -> 5 circles including inner

% Outer radius of the last circle
\pgfmathsetmacro{\rOuter}{\rInner + \nRings*\ringWidth}

% --- Night sector (half annulus) ---
% (removed; we are making a true half-plot instead)
% \fill[gray!70]
%   (0:\rInner)   arc[start angle=0,   end angle=180, radius=\rInner]
%   -- (180:\rOuter) arc[start angle=180, end angle=0,   radius=\rOuter]
%   -- cycle;

% ==========================================================
% Ring-aware cell fill:
% ring 1..4 = W=8,12,15,20 (inner -> outer)
% ==========================================================
\newcommand{\FillCellRing}[4]{%
  % #1 = start hour, #2 = end hour, #3 = ring index 1..4, #4 = color
  \pgfmathsetmacro{\startAng}{90 - 15*#1}
  \pgfmathsetmacro{\endAng}{90 - 15*#2}
  \pgfmathsetmacro{\rA}{\rInner + (#3-1)*\ringWidth}
  \pgfmathsetmacro{\rB}{\rInner + (#3)*\ringWidth}
  \fill[#4, opacity=0.75]
    (\startAng:\rA) -- (\startAng:\rB)
      arc[start angle=\startAng, end angle=\endAng, radius=\rB]
    -- (\endAng:\rA)
      arc[start angle=\endAng, end angle=\startAng, radius=\rA]
    -- cycle;
}

\newcommand{\FillCellAllW}[6]{%
  \FillCellRing{#1}{#2}{1}{#3}%
  \FillCellRing{#1}{#2}{2}{#4}%
  \FillCellRing{#1}{#2}{3}{#5}%
  \FillCellRing{#1}{#2}{4}{#6}%
}

% ==========================================================
% Apply Width-lux-* results (ΔL mapping in comments)
% ==========================================================

% ΔL=5:  W8=RED, W12=RED, W15=RED, W20=YELLOW
\FillCellAllW{6}{7}{red!70}{red!70}{red!70}{yellow!80}
\FillCellAllW{17}{18}{red!70}{red!70}{red!70}{yellow!80}

% ΔL=10: W8=RED, W12=YELLOW, W15=GREEN, W20=GREEN
\FillCellAllW{7}{8}{red!70}{yellow!80}{green!60}{green!60}
\FillCellAllW{16}{17}{red!70}{yellow!80}{green!60}{green!60}

% ΔL=15 (all GREEN)
\FillCellAllW{8}{9}{green!60}{green!60}{green!60}{green!60}
\FillCellAllW{15}{16}{green!60}{green!60}{green!60}{green!60}

% ΔL=20 (all GREEN)
\FillCellAllW{9}{10}{green!60}{green!60}{green!60}{green!60}
\FillCellAllW{14}{15}{green!60}{green!60}{green!60}{green!60}

% ΔL=25 (all GREEN)
\FillCellAllW{10}{11}{green!60}{green!60}{green!60}{green!60}
\FillCellAllW{13}{14}{green!60}{green!60}{green!60}{green!60}

% ΔL=30 (all GREEN)
\FillCellAllW{11}{12}{green!60}{green!60}{green!60}{green!60}

% ΔL=35 (all GREEN)
\FillCellAllW{12}{13}{green!60}{green!60}{green!60}{green!60}

% --- 5 nested *half* circles (semi-doughnut) ---
\foreach \k in {0,...,\nRings} {
  \pgfmathsetmacro{\rr}{\rInner + \k*\ringWidth}
  \draw[thick] (0:\rr) arc[start angle=0, end angle=-180, radius=\rr];
}
% close the annulus edges on the diameter
\draw[thick] (0:\rInner) -- (0:\rOuter);
\draw[thick] (-180:\rInner) -- (-180:\rOuter);

% --- W labels on the 12:00 hour line (inner->outer: 8,12,15,20) ---
\pgfmathsetmacro{\angTwelve}{90 - 15*12}
\foreach[count=\i from 1] \w in {8,12,15,20}{%
  \pgfmathsetmacro{\rMid}{\rInner + (\i-0.5)*\ringWidth}
  \node[font=\footnotesize] at (\angTwelve:\rMid) {$W=\w\,\mathrm{cm}$};
}

% --- Hour rays + labels for the half (6..18) ---
\foreach \h in {6,...,18} {
  \pgfmathsetmacro{\ang}{90 - 15*\h}
  \draw[gray!60] (\ang:\rInner) -- (\ang:{\rOuter+0.10});
  \node[font=\tiny] at (\ang:{\rOuter+0.18}) {\h};
}

% --- Radius for the dashed double-headed arrows ---
\pgfmathsetmacro{\rArrow}{\rOuter + 0.35}

% ==========================================================
% ΔL arc labels along dashed arc
% ==========================================================
\newcommand{\DeltaLArc}[3]{%
  % #1 = ΔL value, #2 = start hour, #3 = end hour
  \pgfmathsetmacro{\startAng}{90 - 15*#2}%
  \pgfmathsetmacro{\endAng}{90 - 15*#3}%
  \draw[<->, dashed, thick]
    (\startAng:\rArrow)
      arc[start angle=\startAng, end angle=\endAng, radius=\rArrow]
      node[midway, sloped, allow upside down=false, above, yshift=0.5pt,
           font=\scriptsize\bfseries, fill=white, inner sep=0.5pt] {$#1$};%
}

% --- ΔL arcs for the day cells ---
\DeltaLArc{5}{6}{7}
\DeltaLArc{10}{7}{8}
\DeltaLArc{15}{8}{9}
\DeltaLArc{20}{9}{10}
\DeltaLArc{25}{10}{11}
\DeltaLArc{30}{11}{12}
\DeltaLArc{30}{12}{13}
\DeltaLArc{25}{13}{14}
\DeltaLArc{20}{14}{15}
\DeltaLArc{15}{15}{16}
\DeltaLArc{10}{16}{17}
\DeltaLArc{5}{17}{18}

\end{tikzpicture}}
    \caption{Width ($W$)}
    \label{fig:ablation_width}
  \end{subfigure}\hfill
  \begin{subfigure}[t]{0.25\textwidth}
    \centering
    \resizebox{\linewidth}{!}{% ---------- DISTANCE (D) ABLATION ----------
\begin{tikzpicture}[scale=2, rotate=180]

% --- Basic radii ---
\def\rInner{0.2}      % inner hole radius
\def\ringWidth{0.3}   % thickness of each ring
\def\nRings{4}        % 4 rings -> 5 circles including inner

% Outer radius of the last circle
\pgfmathsetmacro{\rOuter}{\rInner + \nRings*\ringWidth}

% ==========================================================
% Ring-aware cell fill:
% ring 1..4 = D=1,8,10,20 (inner -> outer)
% ==========================================================
\newcommand{\FillCellRing}[4]{%
  % #1 = start hour, #2 = end hour, #3 = ring index 1..4, #4 = color
  \pgfmathsetmacro{\startAng}{90 - 15*#1}
  \pgfmathsetmacro{\endAng}{90 - 15*#2}
  \pgfmathsetmacro{\rA}{\rInner + (#3-1)*\ringWidth}
  \pgfmathsetmacro{\rB}{\rInner + (#3)*\ringWidth}
  \fill[#4, opacity=0.75]
    (\startAng:\rA) -- (\startAng:\rB)
      arc[start angle=\startAng, end angle=\endAng, radius=\rB]
    -- (\endAng:\rA)
      arc[start angle=\endAng, end angle=\startAng, radius=\rA]
    -- cycle;
}

\newcommand{\FillCellAllD}[6]{%
  % #1,#2 = start/end hour
  % #3..#6 = colors for D=1, D=8, D=10, D=20 (inner->outer)
  \FillCellRing{#1}{#2}{1}{#3}%
  \FillCellRing{#1}{#2}{2}{#4}%
  \FillCellRing{#1}{#2}{3}{#5}%
  \FillCellRing{#1}{#2}{4}{#6}%
}

% ==========================================================
% Fills from Distance-lux-* table (kept in comments)
% ==========================================================

% ΔL=5:  all YELLOW (D1, D8, D10, D20)
\FillCellAllD{6}{7}{yellow!80}{yellow!80}{yellow!80}{yellow!80}
\FillCellAllD{17}{18}{yellow!80}{yellow!80}{yellow!80}{yellow!80}

% ΔL=10..35: all GREEN
\FillCellAllD{7}{8}{green!60}{green!60}{green!60}{green!60}
\FillCellAllD{8}{9}{green!60}{green!60}{green!60}{green!60}
\FillCellAllD{9}{10}{green!60}{green!60}{green!60}{green!60}
\FillCellAllD{10}{11}{green!60}{green!60}{green!60}{green!60}
\FillCellAllD{11}{12}{green!60}{green!60}{green!60}{green!60}
\FillCellAllD{12}{13}{green!60}{green!60}{green!60}{green!60}
\FillCellAllD{13}{14}{green!60}{green!60}{green!60}{green!60}
\FillCellAllD{14}{15}{green!60}{green!60}{green!60}{green!60}
\FillCellAllD{15}{16}{green!60}{green!60}{green!60}{green!60}
\FillCellAllD{16}{17}{green!60}{green!60}{green!60}{green!60}

% ==========================================================
% Gray cells (night) -- removed (we use a half-plot instead)
% ==========================================================
% \fill[gray!70]
%   (0:\rInner)   arc[start angle=0,   end angle=180, radius=\rInner]
%   -- (180:\rOuter) arc[start angle=180, end angle=0,   radius=\rOuter]
%   -- cycle;

% --- 5 nested *half* circles (semi-doughnut) ---
\foreach \k in {0,...,\nRings} {
  \pgfmathsetmacro{\rr}{\rInner + \k*\ringWidth}
  \draw[thick] (0:\rr) arc[start angle=0, end angle=-180, radius=\rr];
}
% close the annulus edges on the diameter
\draw[thick] (0:\rInner) -- (0:\rOuter);
\draw[thick] (-180:\rInner) -- (-180:\rOuter);

% --- D labels on the 12:00 hour line (inner->outer: 1,8,10,20) ---
\pgfmathsetmacro{\angTwelve}{90 - 15*12}
\foreach[count=\i from 1] \d in {1,8,10,20}{%
  \pgfmathsetmacro{\rMid}{\rInner + (\i-0.5)*\ringWidth}
  \node[font=\footnotesize] at (\angTwelve:\rMid) {$D=\d\,\mathrm{cm}$};
}

% --- Hour rays + labels for the half (6..18) ---
\foreach \h in {6,...,18} {
  \pgfmathsetmacro{\ang}{90 - 15*\h}
  \draw[gray!60] (\ang:\rInner) -- (\ang:{\rOuter+0.10});
  \node[font=\tiny] at (\ang:{\rOuter+0.18}) {\h};
}

% --- Radius for the dashed double-headed arrows ---
\pgfmathsetmacro{\rArrow}{\rOuter + 0.35}

% ==========================================================
% ΔL arc labels
% ==========================================================
\newcommand{\DeltaLArc}[3]{%
  % #1 = ΔL value, #2 = start hour, #3 = end hour
  \pgfmathsetmacro{\startAng}{90 - 15*#2}%
  \pgfmathsetmacro{\endAng}{90 - 15*#3}%
  \draw[<->, dashed, thick]
    (\startAng:\rArrow)
      arc[start angle=\startAng, end angle=\endAng, radius=\rArrow]
      node[midway, sloped, allow upside down=false, above, yshift=0.5pt,
           font=\scriptsize\bfseries, fill=white, inner sep=0.5pt] {$#1$};%
}

% --- ΔL arcs ---
\DeltaLArc{5}{6}{7}
\DeltaLArc{10}{7}{8}
\DeltaLArc{15}{8}{9}
\DeltaLArc{20}{9}{10}
\DeltaLArc{25}{10}{11}
\DeltaLArc{30}{11}{12}
\DeltaLArc{30}{12}{13}
\DeltaLArc{25}{13}{14}
\DeltaLArc{20}{14}{15}
\DeltaLArc{15}{15}{16}
\DeltaLArc{10}{16}{17}
\DeltaLArc{5}{17}{18}

\end{tikzpicture}}
    \caption{Distance ($D$)}
    \label{fig:ablation_distance}
  \end{subfigure}\hfill
  \begin{subfigure}[t]{0.25\textwidth}
    \centering
    \resizebox{\linewidth}{!}{% ---------- ANGLE (beta) ABLATION ----------
\begin{tikzpicture}[scale=2, rotate=180]

% --- Basic radii ---
\def\rInner{0.2}      % inner hole radius
\def\ringWidth{0.3}   % thickness of each ring
\def\nRings{4}        % 4 rings -> 5 circles including inner

% Outer radius of the last circle
\pgfmathsetmacro{\rOuter}{\rInner + \nRings*\ringWidth}

% ==========================================================
% Ring-aware cell fill:
% ring 1..4 = \beta=0,4,8,12 (inner -> outer)
% ==========================================================
\newcommand{\FillCellRing}[4]{%
  % #1 = start hour, #2 = end hour, #3 = ring index 1..4, #4 = color
  \pgfmathsetmacro{\startAng}{90 - 15*#1}
  \pgfmathsetmacro{\endAng}{90 - 15*#2}
  \pgfmathsetmacro{\rA}{\rInner + (#3-1)*\ringWidth}
  \pgfmathsetmacro{\rB}{\rInner + (#3)*\ringWidth}
  \fill[#4, opacity=0.75]
    (\startAng:\rA) -- (\startAng:\rB)
      arc[start angle=\startAng, end angle=\endAng, radius=\rB]
    -- (\endAng:\rA)
      arc[start angle=\endAng, end angle=\startAng, radius=\rA]
    -- cycle;
}

\newcommand{\FillCellAllBeta}[6]{%
  % #1,#2 = start/end hour
  % #3..#6 = colors for \beta=0, \beta=4, \beta=8, \beta=12 (inner->outer)
  \FillCellRing{#1}{#2}{1}{#3}%
  \FillCellRing{#1}{#2}{2}{#4}%
  \FillCellRing{#1}{#2}{3}{#5}%
  \FillCellRing{#1}{#2}{4}{#6}%
}

% ==========================================================
% Fills from Angle-lux-* table (kept as-is in comments)
% 06-07: ΔL=5, 07-08: ΔL=10, ..., 17-18: ΔL=5
% ==========================================================

% ΔL=5:  all YELLOW (\beta=0,4,8,12)
\FillCellAllBeta{6}{7}{yellow!80}{yellow!80}{yellow!80}{yellow!80}
\FillCellAllBeta{17}{18}{yellow!80}{yellow!80}{yellow!80}{yellow!80}

% ΔL=10..35: all GREEN
\FillCellAllBeta{7}{8}{green!60}{green!60}{green!60}{green!60}
\FillCellAllBeta{8}{9}{green!60}{green!60}{green!60}{green!60}
\FillCellAllBeta{9}{10}{green!60}{green!60}{green!60}{green!60}
\FillCellAllBeta{10}{11}{green!60}{green!60}{green!60}{green!60}
\FillCellAllBeta{11}{12}{green!60}{green!60}{green!60}{green!60}
\FillCellAllBeta{12}{13}{green!60}{green!60}{green!60}{green!60}
\FillCellAllBeta{13}{14}{green!60}{green!60}{green!60}{green!60}
\FillCellAllBeta{14}{15}{green!60}{green!60}{green!60}{green!60}
\FillCellAllBeta{15}{16}{green!60}{green!60}{green!60}{green!60}
\FillCellAllBeta{16}{17}{green!60}{green!60}{green!60}{green!60}

% ==========================================================
% Gray cells (night) -- removed (we use a half-plot instead)
% ==========================================================
% \fill[gray!70]
%   (0:\rInner)   arc[start angle=0,   end angle=180, radius=\rInner]
%   -- (180:\rOuter) arc[start angle=180, end angle=0,   radius=\rOuter]
%   -- cycle;

% --- 5 nested *half* circles (semi-doughnut) ---
\foreach \k in {0,...,\nRings} {
  \pgfmathsetmacro{\rr}{\rInner + \k*\ringWidth}
  \draw[thick] (0:\rr) arc[start angle=0, end angle=-180, radius=\rr];
}
% close the annulus edges on the diameter
\draw[thick] (0:\rInner) -- (0:\rOuter);
\draw[thick] (-180:\rInner) -- (-180:\rOuter);

% --- \beta labels on the 12:00 hour line (inner->outer: 0,4,8,12) ---
\pgfmathsetmacro{\angTwelve}{90 - 15*12}
\foreach[count=\i from 1] \b in {0,4,8,12}{%
  \pgfmathsetmacro{\rMid}{\rInner + (\i-0.5)*\ringWidth}
  \node[font=\footnotesize] at (\angTwelve:\rMid) {$\beta=\b^\circ$};
}

% --- Hour rays + labels for the half (6..18) ---
\foreach \h in {6,...,18} {
  \pgfmathsetmacro{\ang}{90 - 15*\h}
  \draw[gray!60] (\ang:\rInner) -- (\ang:{\rOuter+0.10});
  \node[font=\tiny] at (\ang:{\rOuter+0.18}) {\h};
}

% --- Radius for the dashed double-headed arrows ---
\pgfmathsetmacro{\rArrow}{\rOuter + 0.35}

% ==========================================================
% ΔL arc labels (diagonal), "ΔL=#" bold
% ==========================================================
\newcommand{\DeltaLArc}[3]{%
  % #1 = ΔL value, #2 = start hour, #3 = end hour
  \pgfmathsetmacro{\startAng}{90 - 15*#2}%
  \pgfmathsetmacro{\endAng}{90 - 15*#3}%
  \draw[<->, dashed, thick]
    (\startAng:\rArrow)
      arc[start angle=\startAng, end angle=\endAng, radius=\rArrow]
      node[midway, sloped, allow upside down=false, above, yshift=0.5pt,
           font=\scriptsize\bfseries, fill=white, inner sep=0.5pt] {$#1$};%
}

% --- ΔL arcs ---
\DeltaLArc{5}{6}{7}
\DeltaLArc{10}{7}{8}
\DeltaLArc{15}{8}{9}
\DeltaLArc{20}{9}{10}
\DeltaLArc{25}{10}{11}
\DeltaLArc{30}{11}{12}
\DeltaLArc{30}{12}{13}
\DeltaLArc{25}{13}{14}
\DeltaLArc{20}{14}{15}
\DeltaLArc{15}{15}{16}
\DeltaLArc{10}{16}{17}
\DeltaLArc{5}{17}{18}

\end{tikzpicture}}
    \caption{Angle ($\beta$)}
    \label{fig:ablation_degree}
  \end{subfigure}
\caption{Illumination-aware ablation of \NSAttack success when varying  (a) length $L$, (b) width $W$, (c) lateral distance $D$, and (d) angle $\beta$. Each subplot varies one geometric parameter while sampling the remaining parameters around the base configuration. Rings represent discrete parameter levels, and angular sectors represent calibrated LAB luminance offsets $\Delta L^* \in \{5,10,15,20,25,30\}$ during daylight hours [06:00,18:00]. Cell color encodes empirical attack success rate: green = high, yellow = intermediate, and red = low.}
  \label{fig:ablation_all_one_row}
\end{figure*}

% \medskip
% Conceptually, the entire loop is governed by two abstract functions. The LD operator
\begin{equation}
  P_{\text{Lane}} = \mathcal{LD}\bigl(I_{\text{rv}}\bigr)
  \label{eq:LD_Function}
\end{equation}
maps any driver-view image to the set of pixels belonging to genuine line markings.  The composite
\begin{equation}
  P_{\text{Lane}}^{\text{NS}}
  \;=\;
  \mathcal{LD}\!\Bigl(
        F_{\text{NS}}\!\bigl(I_{\text{rv}}, W, D, L, \beta\bigr)
      \Bigr)
  \;=\;
  P_{\text{Lane}} \cup P_{\text{NS}}
  \label{eq:F_NS_Function}
\end{equation}

\noindent shows that applying \(F_{\text{NS}}\) casts the \(P_{\text{NS}}\) into the model’s output when a mis-detection occurs. This evaluation loop implements the mappings in \autoref{eq:LD_Function}--\autoref{eq:F_NS_Function} by casting parameterized \NS patterns and measuring whether they are falsely recognized as line markings, thereby determining the conditions that lead to misdetection.~\autoref{fig:syseval_pc} illustrates the parameter-space ranges under which each LD model is consistently fooled by \NSAttack. CLRerNet and TwinLiteNet display broad vulnerability across a span of lateral distances from the line. In contrast, HybridNets tends to be misled primarily by \NSes placed very close to the center line marking or the curb. All three models are consistently affected by \NSes whose width is comparable to that of genuine line markings (roughly 6--36\,cm), with angles (under 15$^\circ$), and lengths exceeding 10\,m.~\autoref{fig:consensus} shows the intersection of successful regions across all three models, identifying a parameter configuration where \NSAttack consistently induces misdetection. This region---characterized by elongated, line-width \NSes positioned near genuine line markings at shallow angles—constitutes a high-confidence vulnerability for vision-based perception systems.

\noindent \textbf{Illumination-Aware Ablation.}
\label{sec:ablation_setup}
We evaluate how \NSAttack success changes as a function of both geometry and illumination. The ablation is centered around a representative configuration inside the consensus region,
$(W_0,L_0,D_0,\beta_0) = (12\,\mathrm{cm},30\,\mathrm{m},10\,\mathrm{cm},8^\circ)$,
which successfully fools all three LD algorithms. Around this base point, we vary one geometric parameter at a time using
$\hat{W} = \{8,12,15,20\}\,\mathrm{cm}$,
$\hat{L} = \{10,20,30,40\}\,\mathrm{m}$,
$\hat{D} = \{1,8,10,20\}\,\mathrm{cm}$, and
$\hat{\beta} = \{0^\circ,4^\circ,8^\circ,12^\circ\}$.

For each geometric setting, we evaluate the attack under six calibrated LAB luminance offsets,
$\Delta L^* \in \{5,10,15,20,25,30\}$,
where larger $\Delta L^*$ values make the \NS brighter relative to the surrounding road. The LAB luminance offset $\Delta L^*$ acts as an approximate brightness-control parameter: we add $\Delta L^*$ to the $L^*$ channel inside the \NS region while preserving the chromatic components. To anchor these values to realistic simulation conditions, we calibrate $\Delta L^*$ empirically against CARLA by matching the rendered \NS--road contrast to representative daytime weather presets. \autoref{fig:illumination_carla} shows one representative rendering for each calibrated illumination level.

In each polar plot in \autoref{fig:ablation_all_one_row}, the concentric rings correspond to discrete values of the ablated geometric parameter, while angular sectors correspond to the calibrated luminance offsets. The color of each cell indicates the empirical attack success rate: green denotes high success, yellow denotes intermediate success, and red denotes low success. Additional ablation details are provided in Appendix~\ref{sec:15-Appendix_E}.

In \autoref{fig:ablation_length}, shorter \NSes with $L \in \{10,20\}\,\mathrm{m}$ become less effective under the darkest calibrated settings, while longer \NSes with $L \in \{30,40\}\,\mathrm{m}$ continue to mislead the LD algorithms across most daylight conditions. \autoref{fig:ablation_width} shows a stronger dependence on width: narrow \NSes with $W \in \{8,12\}\,\mathrm{cm}$ often fail under low luminance contrast, whereas wider \NSes with $W \in \{15,20\}\,\mathrm{cm}$ remain more effective. In contrast, \autoref{fig:ablation_distance} and \autoref{fig:ablation_degree} show smaller variations within the tested consensus range. Overall, once the \NS is sufficiently long and wide, it remains effective across a broad range of calibrated daylight conditions; the main limitations arise from very short or narrow \NS geometries rather than moderate changes in distance or angle.

\section{Safety Impact Analysis}
\label{sec:7-Safety_Evaluation}
In this section, we evaluate the practicality of the \NegativeShadow attack in three different approaches: through software-in-the-loop simulation, a miniature road setup, and real-world testing.

\subsection{Software-in-the-Loop Evaluation}

\label{sec:7_1_Software-in-the-Loop}

We use Openpilot v0.9.5~\cite{commaaiopenpilot2023}, an open source L2 advanced driver assistance system, as an AV stack with CARLA in the loop to evaluate the three defined scenarios. Openpilot hardware, such as the Comma 3X~\cite{comma3x}, integrates with a vehicle's in-vehicle network to automate specific driving tasks, such as steering and acceleration. By leveraging the bridge between Openpilot and CARLA, we can simulate and assess the outcomes of our scenarios in a detailed virtual environment, ensuring thorough testing of Openpilot's decision-making algorithms under challenging conditions. This approach provides valuable insight into the performance and safety of the system in scenarios that mimic real-world situations.

\begin{figure}[!t]
\centering
% Row 1 — Scenario 1
\begin{subfigure}[b]{0.48\columnwidth}
  \centering
  \includegraphics[width=\linewidth,height=1.5cm]{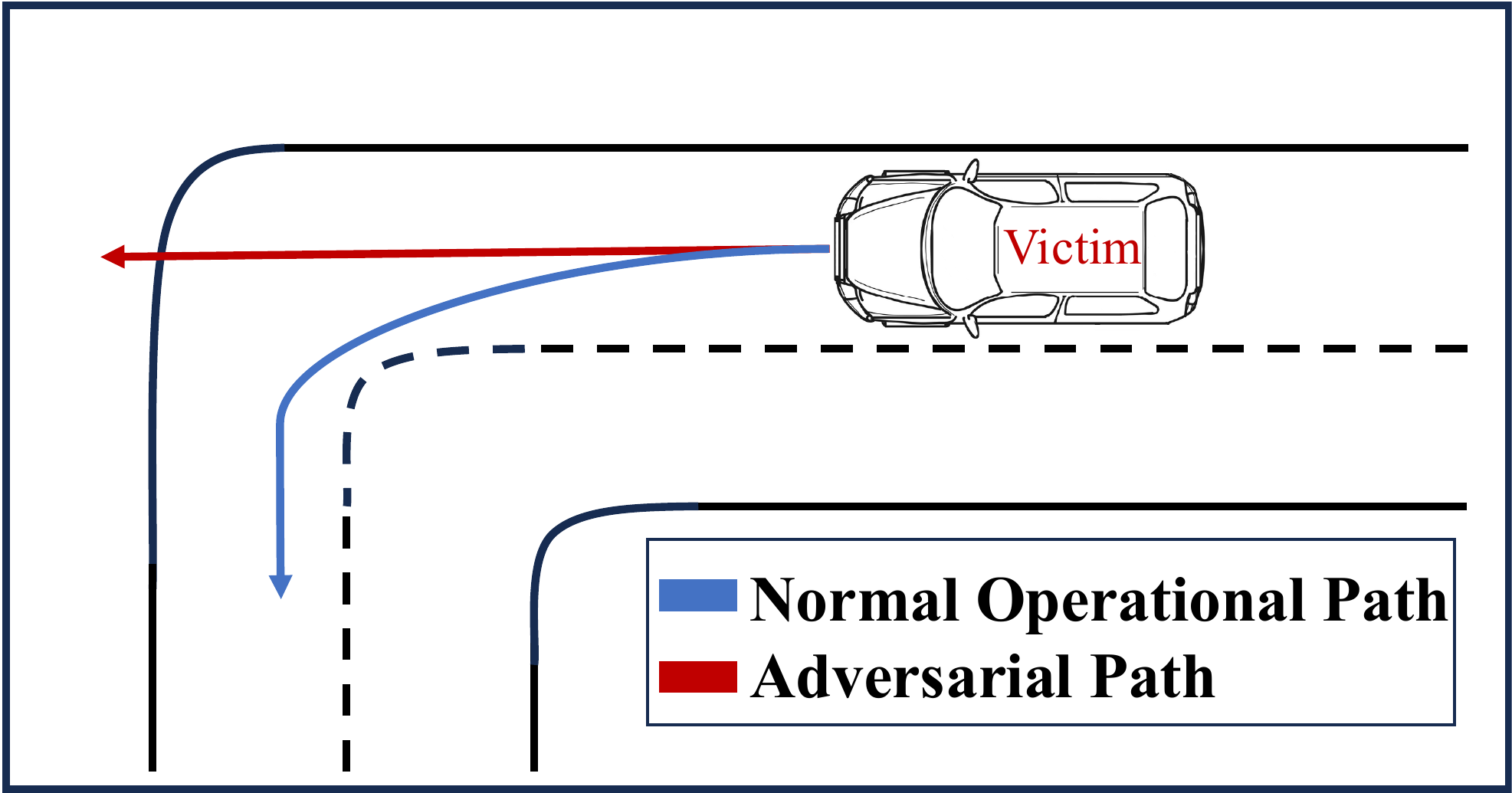}
  \caption{Scenario 1 — Concept\label{fig:first_scenario_concept}}
\end{subfigure}
\hfill
\begin{subfigure}[b]{0.48\columnwidth}
  \centering
  \includegraphics[width=\linewidth,height=1.5cm]{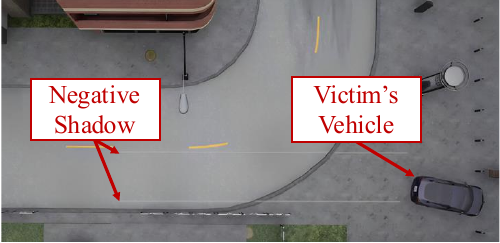}
  \caption{Scenario 1 — Simulation\label{fig:first_scenario}}
\end{subfigure}
\par\medskip
% Row 2 — Scenario 2
\begin{subfigure}[b]{0.48\columnwidth}
  \centering
  \includegraphics[width=\linewidth,height=1.5cm]{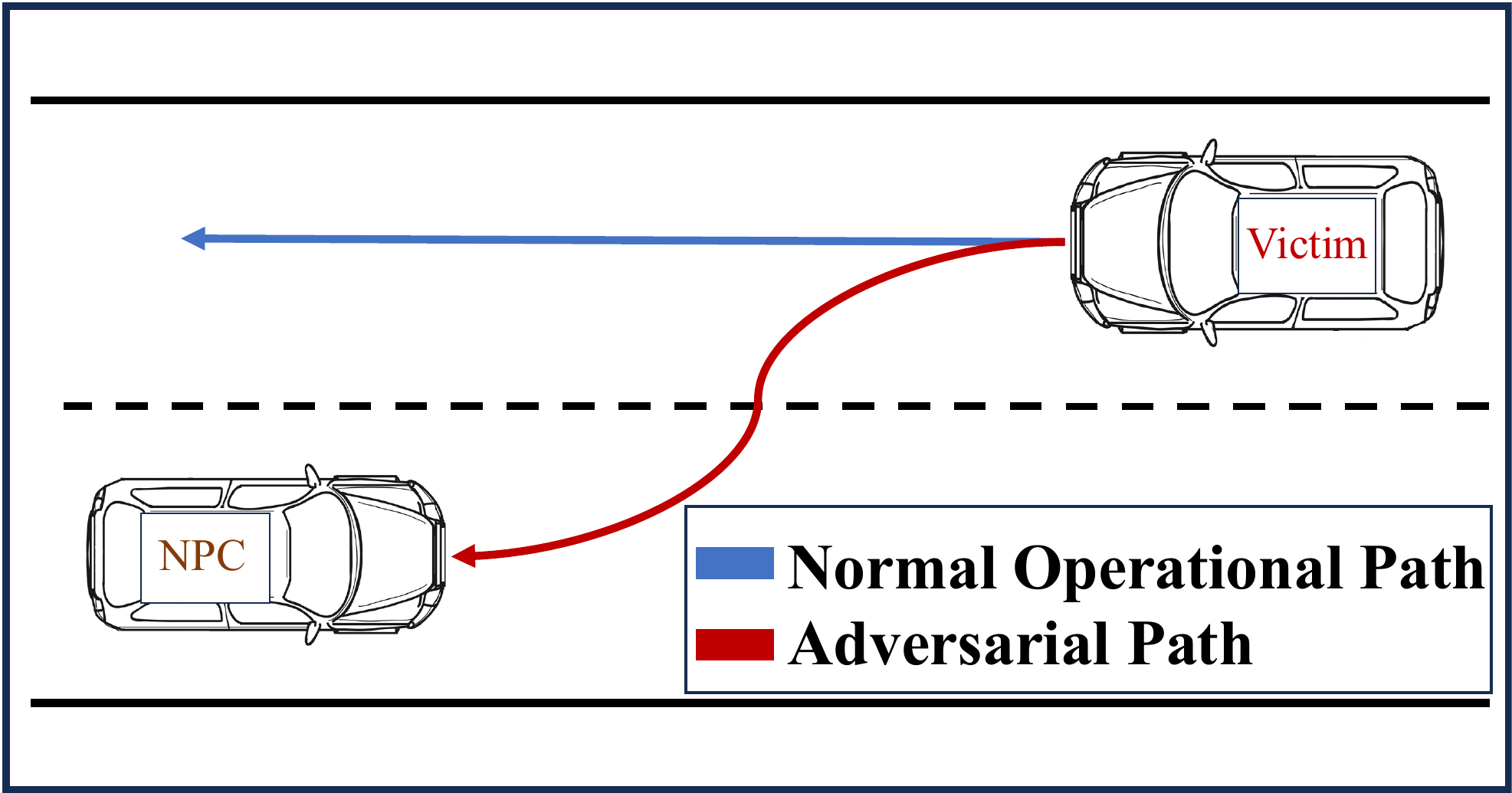}
  \caption{Scenario 2 — Concept\label{fig:second_scenario_concept}}
\end{subfigure}
\hfill
\begin{subfigure}[b]{0.48\columnwidth}
  \centering
  \includegraphics[width=\linewidth,height=1.5cm]{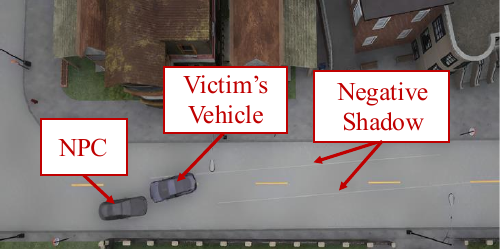}
  \caption{Scenario 2 — Simulation\label{fig:second_scenario}}
\end{subfigure}
\par\medskip
% Row 3 — Scenario 3
\begin{subfigure}[b]{0.48\columnwidth}
  \centering
  \includegraphics[width=\linewidth,height=1.5cm]{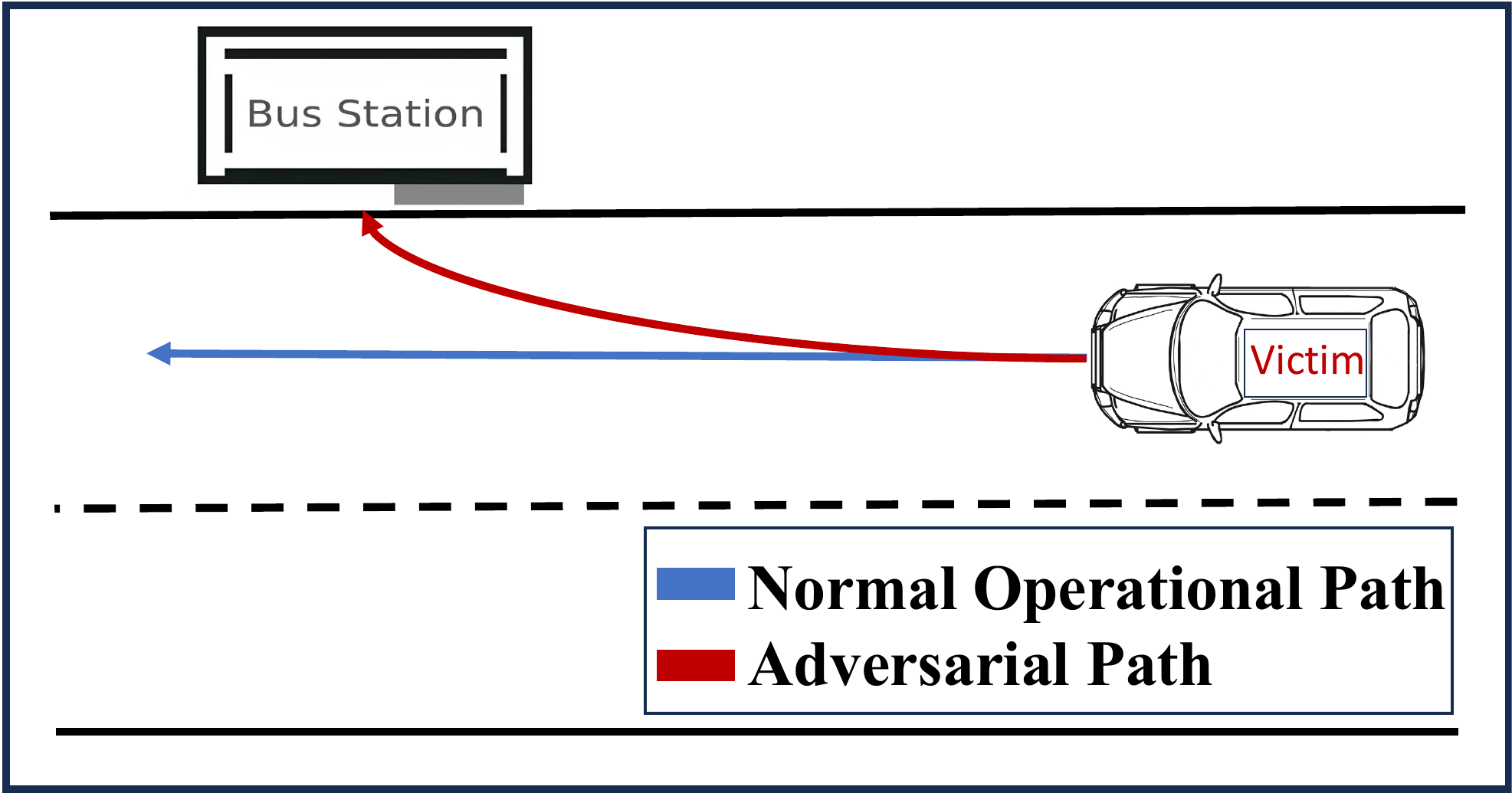}
  \caption{Scenario 3 — Concept\label{fig:third_scenario_concept}}
\end{subfigure}
\hfill
\begin{subfigure}[b]{0.48\columnwidth}
  \centering
  \includegraphics[width=\linewidth,height=1.5cm]{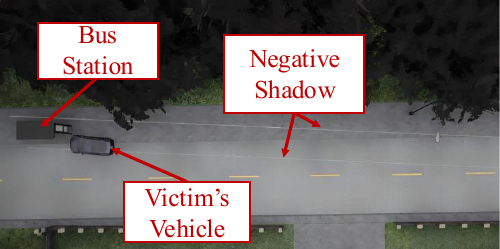}
  \caption{Scenario 3 — Simulation\label{fig:third_scenario}}
\end{subfigure}
\caption{Safety violation scenarios --- concepts and simulations
(red: adversarial, blue: benign).}
\label{fig:safety_eval}
\vspace{-4mm}
\end{figure}

The \textbf{first} scenario, illustrated in \autoref{fig:first_scenario} using the CARLA simulator, shows the car being led off its intended path during a left-turn maneuver, resulting in off-road hazards and collisions. The vehicle continues to use the curb as a reference and correctly initiates the turn. However, it ultimately fails to complete the maneuver and continues straight, veering off the drivable surface. In the \textbf{second} scenario (\autoref{fig:second_scenario}), the car is tricked into a head-on collision with an NPC vehicle. As it drives along a two-way street, the vehicle gradually drifts out of its lane, misinterpreting the opposing lane as part of the valid drivable space. This misdetection leads it to collide directly with the oncoming vehicle. The \textbf{third} scenario in \autoref{fig:third_scenario} shows the victim's car misled into deviating toward roadside infrastructure. The vehicle begins aligned with the genuine lane but eventually veers toward the curb, entering the shoulder area and colliding with a bus station.

Across all three scenarios, a single \NS---though misclassified as a line marking—was insufficient to cause a safety violation. The vehicle interprets the space between the \NS and either the curb or the genuine line marking as drivable. This misdetection is not enough to disrupt the vehicle’s trajectory. As the vehicle proceeds, the AV re-evaluates its environment and falls back on the original lane geometry or the curb to update its drivable space estimation. To overcome this, a second \NS was introduced parallel to and similar in appearance to the first, thereby enclosing the drivable space between the two NSes and leading the AV to interpret them as the new left and right boundaries of a legitimate lane. As a result, the vehicle abandons its reliance on the curb or original line markings.

%%%%%%%%

% \begin{figure*}[!t]
% \centering

% %  Column 1 
% \begin{minipage}{0.22\textwidth}\centering
%   \subfloat[Scenario 1 — Concept\label{fig:first_scenario_concept}]{%
%     \includegraphics[width=\linewidth]{Images/Scenario/First.pdf}}\\[\smallskipamount]
%   \subfloat[Scenario 1 — Simulation\label{fig:first_scenario}]{%
%     \includegraphics[width=\linewidth]{Images/Scenario/S1.pdf}}
% \end{minipage}\hfill
% %  Column 2 
% \begin{minipage}{0.22\textwidth}\centering
%   \subfloat[Scenario 2 — Concept\label{fig:second_scenario_concept}]{%
%     \includegraphics[width=\linewidth]{Images/Scenario/Second.pdf}}\\[\smallskipamount]
%   \subfloat[Scenario 2 — Simulation\label{fig:second_scenario}]{%
%     \includegraphics[width=\linewidth]{Images/Scenario/S2.pdf}}
% \end{minipage}\hfill
% %  Column 3 
% \begin{minipage}{0.22\textwidth}\centering
%   \subfloat[Scenario 3 — Concept\label{fig:third_scenario_concept}]{%
%     \includegraphics[width=\linewidth]{Images/Scenario/Third.pdf}}\\[\smallskipamount]
%   \subfloat[Scenario 3 — Simulation\label{fig:third_scenario}]{%
%     \includegraphics[width=\linewidth]{Images/Scenario/S3.pdf}}
% \end{minipage}\hfill
% %  Column 4 
% \begin{minipage}{0.22\textwidth}\centering
%   \subfloat[Miniature road testbed\label{fig:miniroad}]{%
%     \includegraphics[width=0.90\linewidth]{Images/Miniature_Road/Mini_Road.pdf}}
% \end{minipage}

% \caption{\centering Safety violation scenarios — concepts and simulations (red: adversarial, blue: benign) and real-world testbed.}
% \label{fig:safety_eval}
% \end{figure*}

\begin{table}[b]
\centering
\scriptsize
\setlength{\tabcolsep}{3pt}
\renewcommand{\arraystretch}{1.0}
\caption{Comparative analysis of simulated attack outcomes on Openpilot in the scenarios}
\label{tab:analysis_table}

\makebox[\columnwidth][c]{%
  \resizebox{0.75\columnwidth}{!}{%
    \begin{tabular}{cccccc}
    \toprule
    \multicolumn{6}{c}{\hspace{1cm}\textbf{Speed (mph)}}\\[3pt]
    \cline{2-6}\\[-1pt]
    \textbf{Length (m)} & \textbf{10} & \textbf{15} & \textbf{20} & \textbf{35} & \textbf{60} \\
    \midrule
    \textbf{25} &
    \textcolor{green!70!black}{\rule{0.7em}{0.7em}}\textcolor{green!70!black}{\rule{0.7em}{0.7em}}\textcolor{red!70!black}{\rule{0.7em}{0.7em}} &
    \textcolor{green!70!black}{\rule{0.7em}{0.7em}}\textcolor{green!70!black}{\rule{0.7em}{0.7em}}\textcolor{red!70!black}{\rule{0.7em}{0.7em}} &
    \textcolor{green!70!black}{\rule{0.7em}{0.7em}}\textcolor{green!70!black}{\rule{0.7em}{0.7em}}\textcolor{red!70!black}{\rule{0.7em}{0.7em}} &
    \textcolor{green!70!black}{\rule{0.7em}{0.7em}}\textcolor{green!70!black}{\rule{0.7em}{0.7em}}\textcolor{green!70!black}{\rule{0.7em}{0.7em}} &
    \textcolor{green!70!black}{\rule{0.7em}{0.7em}}\textcolor{red!70!black}{\rule{0.7em}{0.7em}}\textcolor{green!70!black}{\rule{0.7em}{0.7em}} \\
    \textbf{30} &
    \textcolor{green!70!black}{\rule{0.7em}{0.7em}}\textcolor{green!70!black}{\rule{0.7em}{0.7em}}\textcolor{red!70!black}{\rule{0.7em}{0.7em}} &
    \textcolor{green!70!black}{\rule{0.7em}{0.7em}}\textcolor{green!70!black}{\rule{0.7em}{0.7em}}\textcolor{red!70!black}{\rule{0.7em}{0.7em}} &
    \textcolor{green!70!black}{\rule{0.7em}{0.7em}}\textcolor{green!70!black}{\rule{0.7em}{0.7em}}\textcolor{green!70!black}{\rule{0.7em}{0.7em}} &
    \textcolor{green!70!black}{\rule{0.7em}{0.7em}}\textcolor{green!70!black}{\rule{0.7em}{0.7em}}\textcolor{green!70!black}{\rule{0.7em}{0.7em}} &
    \textcolor{green!70!black}{\rule{0.7em}{0.7em}}\textcolor{red!70!black}{\rule{0.7em}{0.7em}}\textcolor{red!70!black}{\rule{0.7em}{0.7em}} \\
    \textbf{35} &
    \textcolor{green!70!black}{\rule{0.7em}{0.7em}}\textcolor{green!70!black}{\rule{0.7em}{0.7em}}\textcolor{red!70!black}{\rule{0.7em}{0.7em}} &
    \textcolor{green!70!black}{\rule{0.7em}{0.7em}}\textcolor{green!70!black}{\rule{0.7em}{0.7em}}\textcolor{green!70!black}{\rule{0.7em}{0.7em}} &
    \textcolor{green!70!black}{\rule{0.7em}{0.7em}}\textcolor{green!70!black}{\rule{0.7em}{0.7em}}\textcolor{green!70!black}{\rule{0.7em}{0.7em}} &
    \textcolor{green!70!black}{\rule{0.7em}{0.7em}}\textcolor{green!70!black}{\rule{0.7em}{0.7em}}\textcolor{green!70!black}{\rule{0.7em}{0.7em}} &
    \textcolor{green!70!black}{\rule{0.7em}{0.7em}}\textcolor{green!70!black}{\rule{0.7em}{0.7em}}\textcolor{green!70!black}{\rule{0.7em}{0.7em}} \\
    \bottomrule
    \end{tabular}%
  }%
}
\vspace{2pt}
{\scriptsize
\textcolor{green!70!black}{\rule{0.6em}{0.6em}} = \textbf{scenario success},\quad
\textcolor{red!70!black}{\rule{0.6em}{0.6em}} = \textbf{scenario failure.}
}
\end{table}

To assess the efficiency and effectiveness of each scenario, we varied \NSes lengths from 25 to 35m and tested speeds from 10 mph to 60 mph (\autoref{tab:analysis_table}). The \textcolor{green!70!black}{\rule{0.7em}{0.7em}} indicates a successful attack, while the \textcolor{red!70!black}{\rule{0.7em}{0.7em}} denotes a failure. Higher vehicle speeds generally require longer \NSes for successful attacks, with the correlation between vehicle speed and \NS length being crucial for determining attack efficacy.

Analysis of the table data reveals diverse risk patterns for each scenario. In the \textbf{first} scenario, the risk of vehicular misdirection increases with the \NS length, regardless of speed. The \textbf{second} scenario indicates a higher risk of collision beyond 30 meters, especially at higher speeds. The \textbf{third} scenario shows a low risk in varying speeds and \NS lengths, except at high speeds (35 and 60 mph) and longer \NS lengths (25 to 35m), where the danger increases. These results reveal that \NS length is the dominant factor driving attack success across all scenarios, with longer shadows consistently leading to misdetection and unsafe  maneuvers---even at lower speeds. As shown in \autoref{fig:success_rate}, the average success rate increases with the length of the \NS in these scenarios, reaching an average of 93.33\% at a \NS length of 35m. This underscores the critical influence of \NS on the success of the scenario; per-speed driver reaction times are reported in Appendix~\ref{sec:12-Appendix_B}. %The reaction time evaluation process and results are provided in \autoref{sec:12-Appendix_B}.

\begin{figure}[b]
\centering
\resizebox{\columnwidth}{!}{%
\begin{tikzpicture}[x=0.06cm,y=0.45cm]
  % miniature road length (cm)
  \def\RoadLen{165}
  % NS front position (cm)
  \def\NSstart{68}
  % NS length from your text: L ≈ 64 cm
  \def\NSlen{64}
  \pgfmathsetmacro{\NSend}{\NSstart+\NSlen}
  % (lux, first-misdetection distance, row index)
  \foreach \lux/\dist/\row in {
    40/64/0,
    90/57/1,
    170/47/2,
    225/40/3,
    330/36/4,
    430/27/5,
    640/18/6,
    750/10/7
  }{
    % full road (0–RoadLen cm) – light background
    \draw[fill=gray!10,draw=gray!40]
      (0,\row-0.3) rectangle (\RoadLen,\row+0.3);
    % NS region (orange band, patterned)
    \draw[
      pattern=crosshatch,
      pattern color=orange!90,
      draw=orange!90
    ]
      (\NSstart,\row-0.3) rectangle (\NSend,\row+0.3);
    % distance to onset of misdetection (blue overlay, patterned)
    \draw[
      pattern=north east lines,
      pattern color=blue!70,
      draw=blue!70
    ]
      (0,\row-0.3) rectangle (\dist,\row+0.3);
    % start point marker
    \fill[black] (0,\row) circle (1pt);
    % onset of misdetection marker
    \fill[red] (\dist,\row) circle (1.5pt);
    % illuminance label
    \node[left] at (-2,\row) {\small \lux\,lx};
  }
  % global vertical dashed lines
  % start (0 cm)  ---> GREEN
  \draw[green!70!black,dashed,thick] (0,-0.7) -- (0,7.4);
  % NS front (\NSstart cm)
  \draw[orange!80!black,dashed,thick] (\NSstart,-0.7) -- (\NSstart,7.4);
  % NS end (\NSend cm)
  \draw[orange!80!black,dashed,thick] (\NSend,-0.7) -- (\NSend,7.4);
  % x-axis ticks and labels
  \foreach \x in {0,25,50,75,100,125,150,165}{
    \draw (\x,-0.6) -- (\x,-0.3);
    \node[below] at (\x,-0.6) {\scriptsize \x};
  }
  \node[below] at (\RoadLen/2,-1.2) {\small Distance from starting point (cm)};
  
  % 2x2 legend
  \begin{scope}[shift={(5,8.5)}, every node/.style={font=\normalsize}]
    \draw[black] (-2,-0.8) rectangle (151,0.8);
    
    % Row 1, Column 1: NS region
    \draw[
      pattern=crosshatch,
      pattern color=orange!90,
      draw=orange!90
    ] (0,0.15) rectangle (3,0.45);
    \node[right] at (3.3,0.35) {NS region};
    
    % Row 1, Column 2: Distance to first misdetection
    \draw[
      pattern=north east lines,
      pattern color=blue!70,
      draw=blue!70
    ] (72,0.15) rectangle (75,0.45);
    \node[right] at (75.3,0.35) {Distance to First Misdetection};
    
    % Row 2, Column 1: Starting Position (GREEN dashed)
    \draw[green!70!black,dashed,line width=3pt] (0,-0.35) -- (4,-0.35);
    \node[right] at (4.3,-0.35) {Starting Position};
    
    % Row 2, Column 2: Start of misdetection
    \fill[red] (75,-0.35) circle (1.5pt);
    \node[right] at (75.7,-0.35) {Start of Misdetection};
  \end{scope}
\end{tikzpicture}%
}
\caption{For each illuminance level, the horizontal bar represents the full 1.65\,m miniature road (0\,cm = starting point, 165\,cm = road end). The blue segment shows how far Comma~3X travels from the starting point before Openpilot first begins to misdetect the \NS as a lane boundary (red dot). The orange segment marks the physical location of the negative shadow along the road. The green dashed line indicates the starting position. Brighter \NSes lead to earlier onset of misdetection (shorter blue segment).}
\label{fig:lux_distance_tracks}
\end{figure}

%%%%%%%%%%%%%%%%%%%%%%%%%%%%%%%%%%%%%%%%%%%%%%%%%%

% \begin{figure}[t]
%   \centering
%   \resizebox{0.95\columnwidth}{!}{%
%     \begin{tikzpicture}
%       \begin{axis}[
%         height=4.5cm,
%         width=9cm,
%         xlabel={Illuminance (lx)},
%         ylabel={Distance to first misdetection (cm)},
%         xmin=30, xmax=780,
%         ymin=0,  ymax=70,
%         xtick={40,140,225,330,430,560,640,750},
%         ytick={0,10,20,30,40,50,60,70},
%         ymajorgrids,
%         xmajorgrids,
%         major grid style={draw=gray!40, dashed},
%         tick label style={font=\footnotesize},
%         label style={font=\footnotesize},
%         axis x line*=bottom,
%         axis y line*=left,
%         x axis line style={->},
%         y axis line style={->},
%         mark size=2pt
%       ]

%         % (lux, distance_cm) pairs
%         \addplot[
%           color=blue!80!black,
%           mark=*,
%           thick
%         ] coordinates {
%           (40,64)
%           (90,57)
%           (140,52)
%           (170,47)
%           (195,45)
%           (225,40)
%           (330,36)
%           (430,27)
%           (560,23)
%           (640,18)
%           (750,10)
%         };

%       \end{axis}
%     \end{tikzpicture}%
%   }
%   \caption{Relationship between illuminance of \NSes and the distance from the starting point at which Openpilot first misdetects them as lane markings.}
%   \label{fig:lux_distance}
% \end{figure}

\subsection{Miniature Testbed Evaluation}
\label{ss:miniature_road}

\begin{figure}[t]
\centering
\begin{minipage}[c]{0.5\linewidth}
  \centering
  \subfloat[Starting Position\label{subfig:Step1_Miniature_Road}]{%
    \includegraphics[width=\linewidth,height=1.5cm]{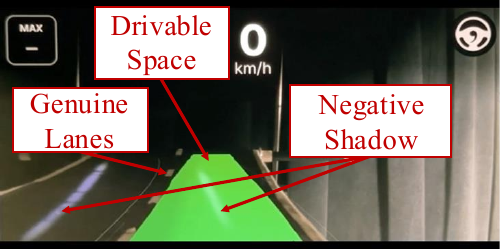}}\\[1mm]
  \subfloat[Midway Point\label{subfig:Step2_Miniature_Road}]{%
    \includegraphics[width=\linewidth,height=1.5cm]{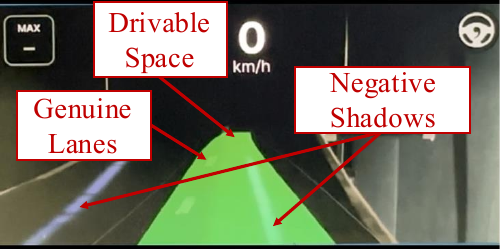}}\\[1mm]
  \subfloat[Trajectory Change\label{subfig:Step3_Miniature_Road}]{%
        \includegraphics[width=\linewidth,height=1.5cm]{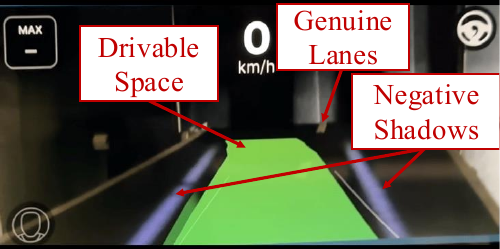}}
\end{minipage}
\hfill
\begin{minipage}[c]{0.48\linewidth}
  \centering
  \subfloat[Miniature road testbed\label{fig:miniroad}]{%
    \includegraphics[width=\linewidth,height=5.8cm]{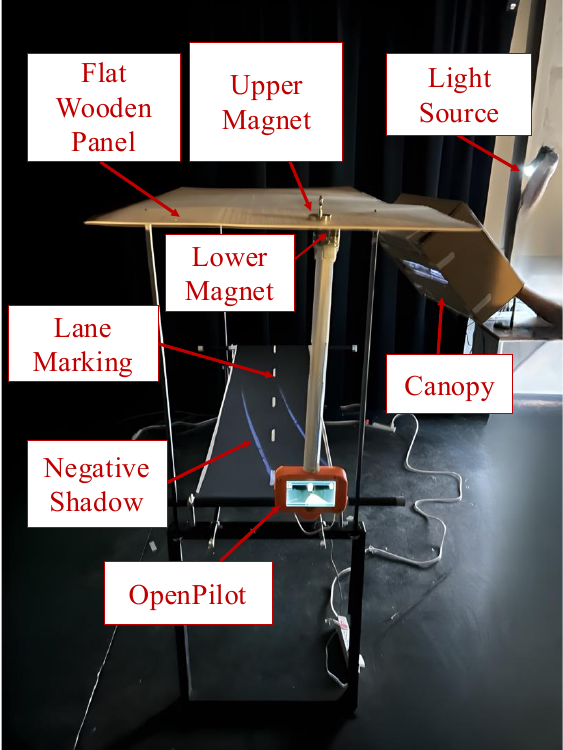}}
\end{minipage}
\vspace{-2mm}
\caption{Evaluation of OpenPilot on a miniature road in the presence of 
negative shadows.}
\label{fig:carla_openpilot}
\vspace{-4mm}
\end{figure}

We evaluate the feasibility of the \NSAttack in a controlled physical setup by implementing the second scenario, \textbf{Head-on Collision}, from Section~\ref{sec:7_1_Software-in-the-Loop}. The goal is to test whether Openpilot misdetects \NSes as lane markings and shifts the perceived drivable space into the opposing lane. To this end, we constructed a miniature road testbed, shown in~\autoref{fig:miniroad}, using scaled lane markings based on typical U.S. road dimensions. Additional details on the miniature road dimensions, scaling procedure, and physical construction are provided in Appendix~\ref{sec:14-Appendix_D}.

A Comma 3X device running Openpilot v0.9.5 was moved along the miniature road using a magnet-based mechanism, while two parallel \NSes were cast onto the road using a canopy with two openings and an external light source. The resulting \NS parameters were within the effective range identified in our earlier analysis: \(D \approx 9\,\mathrm{cm}\), \(\beta \approx 5^\circ\), \(W \approx 2.2\,\mathrm{cm}\), and \(L \approx 64\,\mathrm{cm}\).

\autoref{subfig:Step1_Miniature_Road} shows the starting position, where Comma 3X correctly detects the genuine lane markings and identifies the drivable space between them. As Openpilot moves forward, it mistakenly identifies the right \NS as a genuine lane marking, shifting the perceived drivable space between the original marking and the right \NS, as shown in \autoref{subfig:Step2_Miniature_Road}. Further along the trajectory, Openpilot overlooks the original marking, detects the left \NS as a genuine lane marking, and relocates the drivable space between the two \NSes, as shown in \autoref{subfig:Step3_Miniature_Road}. This causes Openpilot to deviate from the original lane markings, confirming the behavior observed in simulation.

We repeated the experiment multiple times to study the relationship between \NS illuminance, measured in lux, and the onset of Openpilot misdetection. In each trial, the Comma 3X was moved forward from a fixed starting point where Openpilot initially detected the genuine lane markings correctly. We then varied the illuminance of the projected \NSes and recorded the distance from the starting point at which Openpilot first misinterpreted the \NSes as lane markings.

As illustrated in \autoref{fig:lux_distance_tracks}, higher illuminance leads to earlier misdetection. At low illuminance, around 40\,lux, Comma~3X travels approximately 63.5\,cm before first interpreting the \NS as a lane boundary. In contrast, at high illuminance, around 757\,lux, misdetection begins as early as 10.16\,cm from the starting point. In other words, brighter \NSes cause Openpilot to stop relying on the genuine lane markings and instead use the \NSes as lane boundaries much earlier along the trajectory.

\begin{figure*}[!t]
\centering
\subfloat[Initial correct lane detection\label{fig:state1}]{%
  \includegraphics[width=0.21\textwidth]{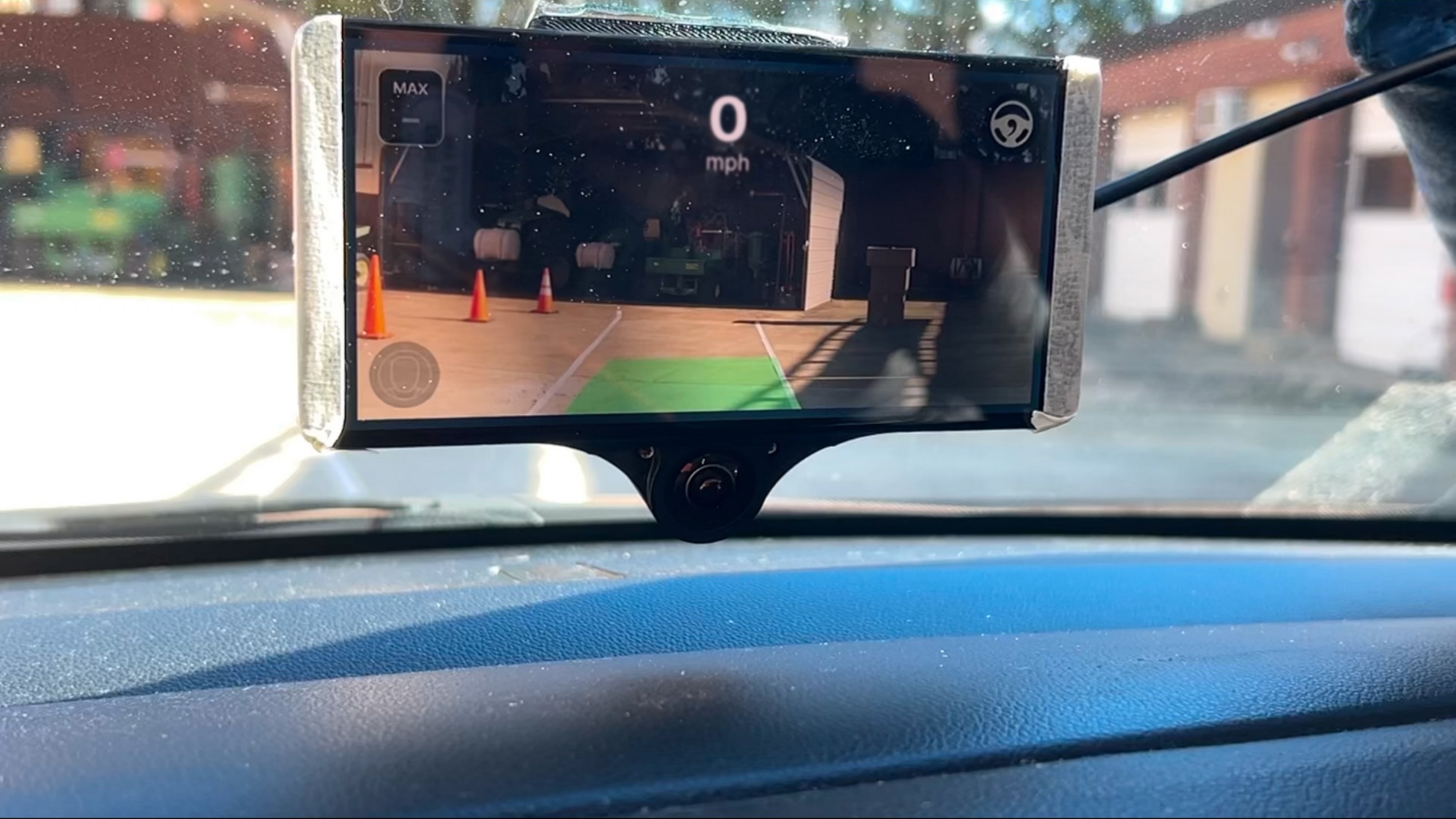}}
\hfill
\subfloat[Left NS misdetected as line\label{fig:state2}]{%
  \includegraphics[width=0.21\textwidth]{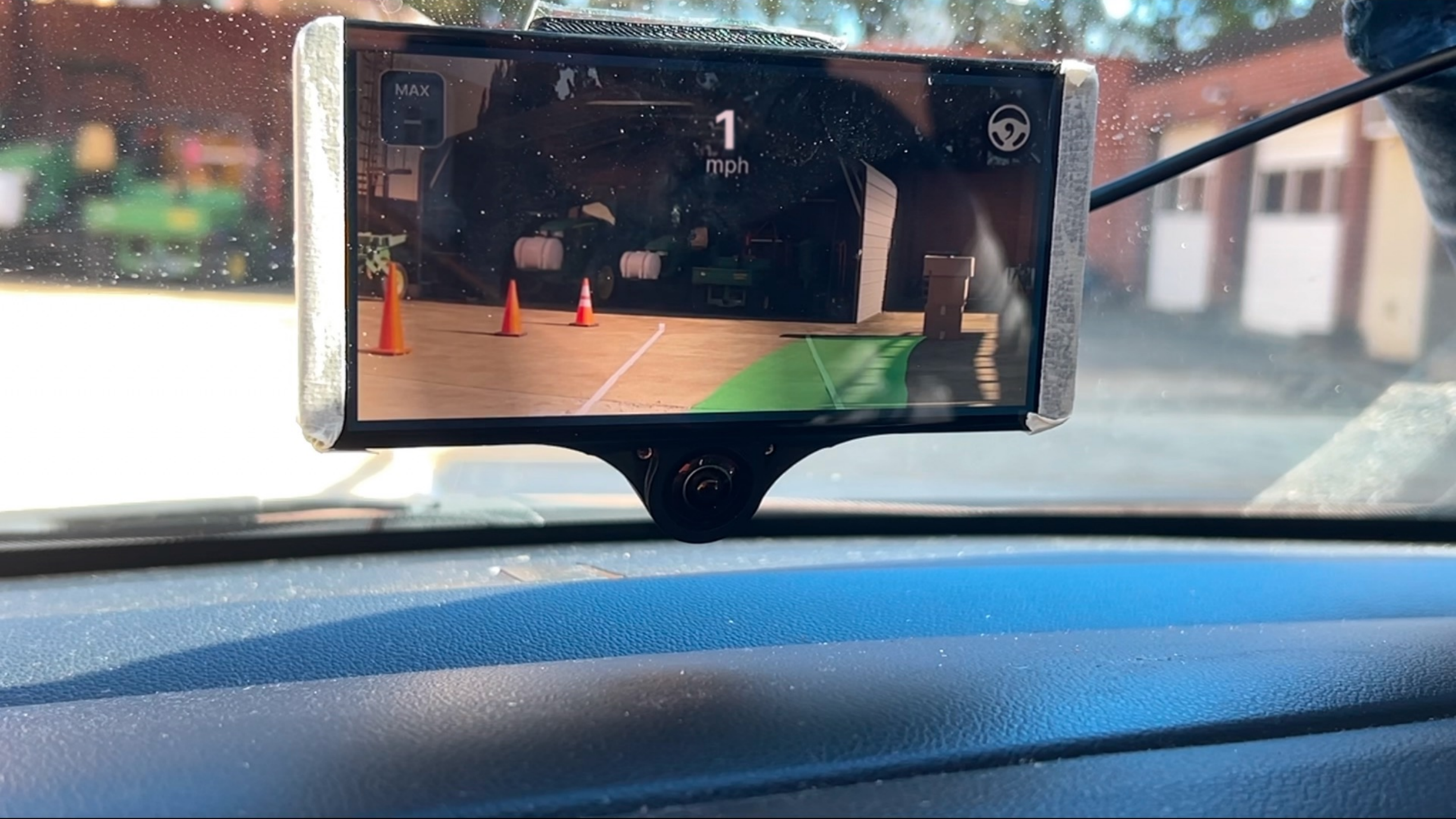}}
\hfill
\subfloat[Drivable space misplaced\label{fig:state3}]{%
  \includegraphics[width=0.21\textwidth]{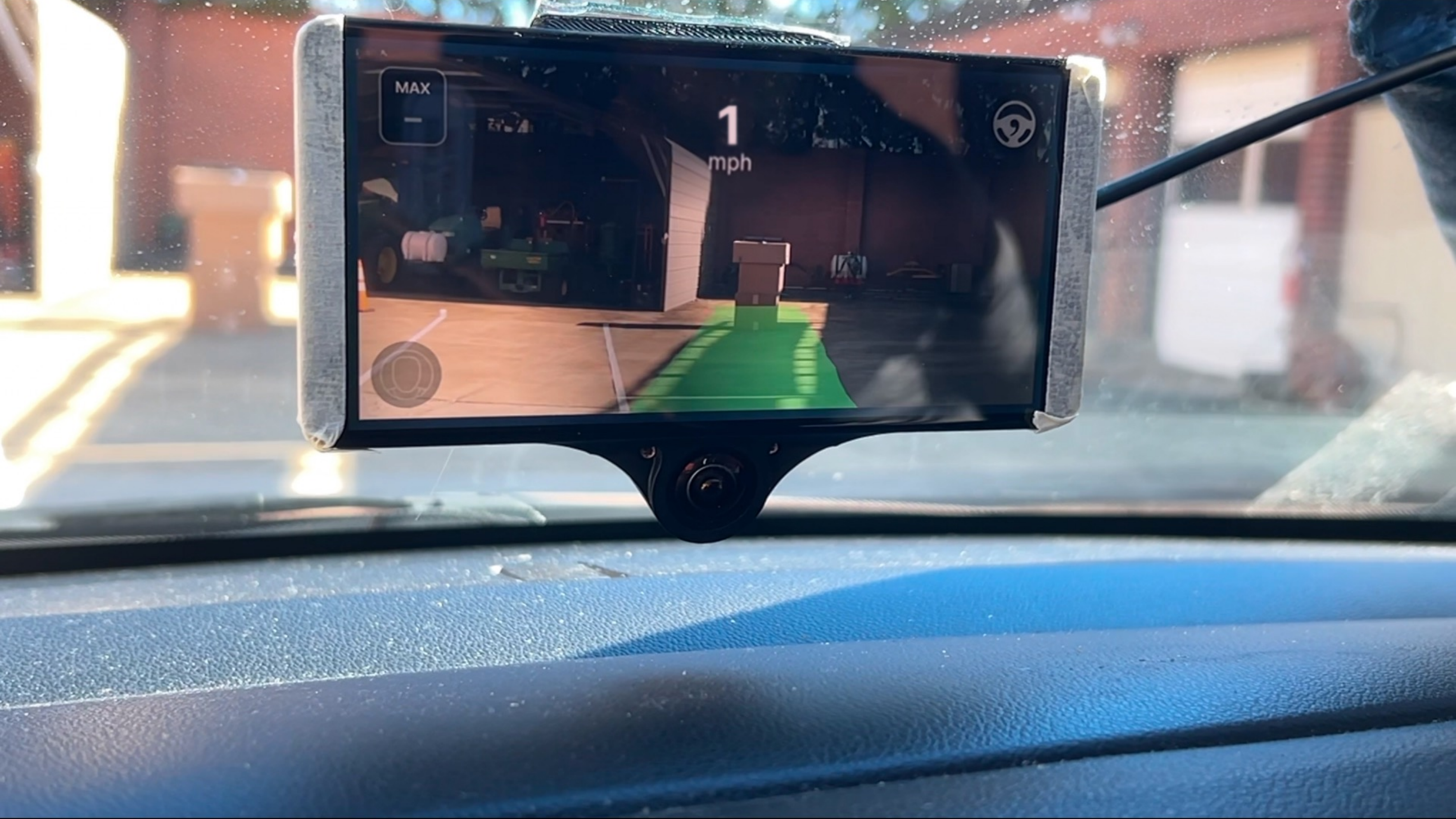}}
\hfill
\subfloat[Collision with barriers\label{fig:state4}]{%
  \includegraphics[width=0.21\textwidth]{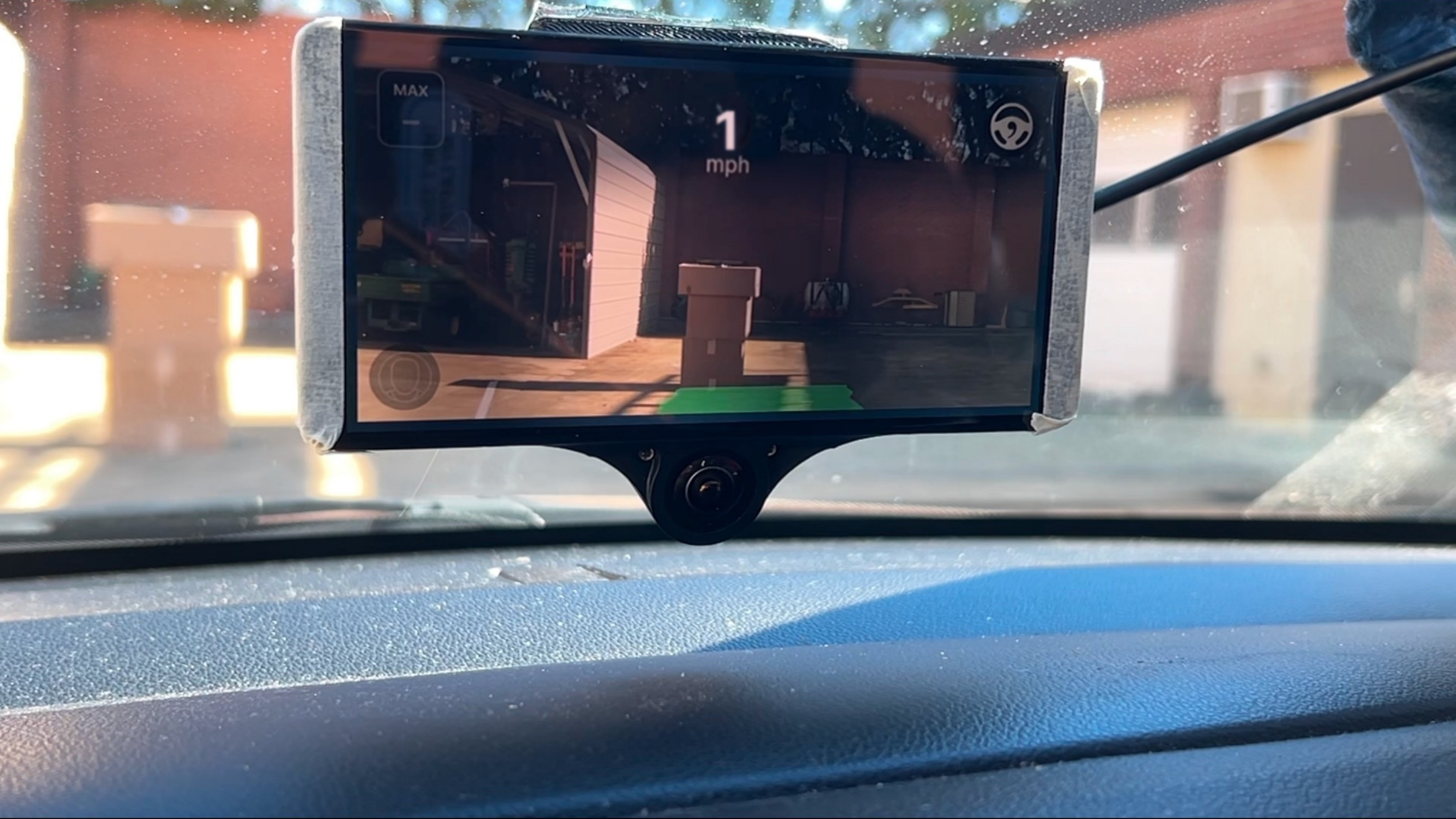}}
\vspace{-1mm}
\caption{Comma 3X misdetected negative shadows as line markings, causing a collision during the real-world test.}
\label{fig:states}
\end{figure*}

\subsection{Real-World Field Evaluation}

\begin{figure}[t]
    \centering
    \includegraphics[width=0.95\columnwidth]{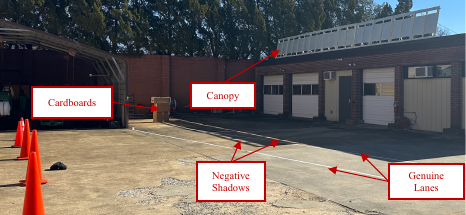}
    \caption{Real-world setup.}
    \label{fig:Real-World-Setup}
    \vspace{-5mm}
\end{figure}

In this subsection, we evaluate the practicality of the \NSAttack within a real-world environment by replicating the third scenario (\textbf{Off-road Deviation}) outlined in Section \ref{sec:7_1_Software-in-the-Loop}. Due to safety considerations and regulatory permissions, the experiment was simulated in a controlled area similar to a street.
% \textbf{Setup Description.} 
We set up two U.S. standard line markings with each 16cm wide~\cite{civilsir2024,FHWA2023PavementMarking}. The markings are separated by a 1.6m-wide gap, representing one lane of a road. Each line extended 20m in length. A canopy was installed 2.95m above a nearby building to project the \NSes onto the road. The canopy dimensions were 12.5m in length and 1.2m in width, featuring two parallel 12cm-wide strips cut out to simulate line markings, as depicted in \autoref{fig:Real-World-Setup}.  %(coordinates REDACTED) 

The shadow length (\(SL\)), equivalent to the \NS length, was calculated using the equations from Section~\ref{sec:5_2_Shadow_Calculation}. Conducted at 10:00 AM within the attack window of 9:00 AM to 4:00 PM during winter months, the shadow parameters stood at \(SL = 18.26 \, \text{m}\), \(D \approx 65 \, \text{cm}\), \(\beta \approx 6^\circ\), \(W \approx 13.7 \, \text{cm}\), and \(L \approx 18.5 \, \text{m}\), all within the successful parameter ranges for effective \NS detection. A Comma 3X device, running Openpilot v0.9.5, was mounted in a 2022 Honda Civic. The car started movement from a designated starting point where genuine line markings were correctly identified, similar to the initial detection phase in the miniature setup (\autoref{fig:state1}).

\begin{enumerate}[noitemsep, nolistsep]
    \item \textbf{First \NS Detection:} The system began to misdetect a \NS as a legitimate line marking, shifting the drivable space between the original line marking and the left \NS (\autoref{fig:state2}).
    \item \textbf{Second \NS Detection:} Further movement caused another \NS on the opposite side to be misidentified, relocating the drivable space between the two \NSes (\autoref{fig:state3}).
    \item \textbf{Final Adjustment:} The Comma 3X maintained the drivable space between the \NSes, steering the vehicle towards pre-set cardboard barriers, resulting in a collision (\autoref{fig:state4}).
\end{enumerate}

This sequence mirrors the misdetection patterns observed in the miniature setup but within a scaled real-world environment, showing the \NSAttack's effectiveness under practical conditions. The entire real-world setup was implemented at a cost of less than \$250, underscoring the affordability and accessibility of executing the \NSAttack in realistic scenarios.
Insights from the miniature road experiments informed the design choices for the real-world testing, particularly in scaling parameters and shadow length calculations. While the miniature setup provided controlled conditions to observe Openpilot's responses, the real-world testing validated these findings in a more dynamic and scalable environment, highlighting the attack's viability beyond simulations.

\section{Human Factors and Stealth Assessment}
\label{sec:8-Human_Factor_Evaluation}

To evaluate the stealthiness of the \NSAttack, we conducted a human-subject study using videos generated in CARLA and Unreal Engine 4~\cite{UnrealEngine2023}. The study was approved by our university's IRB. We selected \texttt{Town04} from CARLA because it provides a realistic town-like environment with a long, straight road suitable for evaluating the proposed attack. To avoid visual confounders, we removed objects with strong illumination or high reflectivity that could interfere with the visibility or alignment of the \NSes.

We designed the attack by positioning an opaque panel with two parallel openings so that sunlight passing through the openings cast bright lane-like patterns along the right side of a straight road. The resulting \NSes were placed next to and parallel to the original lane markings, mimicking lane markings on one side of the street.

The study evaluated participant responses under two weather conditions: a clear day and a bright overcast setting with recently wet ground. These weather conditions were selected because illumination and road-surface reflectance can affect the visibility and perceived naturalness of the \NSes. For each weather condition, we generated both an attack scenario and a corresponding control scenario without the attack. The control videos allowed us to compare participant responses under normal driving conditions against their responses when \NSes were present.

We used CARLA's built-in autonomous driving functionality in cautious mode to generate the vehicle trajectories in the prerecorded videos. The mean vehicle speeds in the clear-day and wet-road attack videos were 2.78 and 2.22\,m/s, respectively; the corresponding control speeds were 3.33 and 2.50\,m/s. During the attack videos, the \NSes appeared only during specific time intervals. Vehicle and pedestrian traffic were excluded from the recordings to ensure that participant attention remained focused on the roadway and the visual effect of the \NSes. Additional details of the study protocol, participant instructions, response logging, exclusion criteria, and participant demographics are provided in Appendix~\ref{sec:13-Appendix_C}.

\begin{figure*}[t]
\centering
\resizebox{0.82\textwidth}{!}{%
\begin{tikzpicture}

\providecolor{olive}{rgb}{0.5,0.5,0}

\pgfplotsset{
    CLRerNetStyle/.style    ={ybar, fill=red!80!black,   draw=none, fill opacity=1},
    HybridNetStyle/.style   ={ybar, fill=blue!80!black,  draw=none, fill opacity=1},
    TwinLiteNetStyle/.style ={ybar, fill=olive!80!black, draw=none, fill opacity=1},
}

\begin{groupplot}[
    group style={
        group size=5 by 1,
        horizontal sep=1.2em
    },
    ybar=1pt,
    /pgf/bar width=2pt,
    ymajorgrids,
    major grid style={draw=gray!30, dashed},
    tick label style={font=\scriptsize\bfseries},
    yticklabel style={xshift=2pt},
    label style={font=\scriptsize\bfseries},
    ylabel style={font=\scriptsize\bfseries},
    xlabel style={font=\scriptsize\bfseries},
    axis line style={line width=0.9pt},
    clip=true,
    width=4cm,
    height=4cm,
]

% ============================================================
% 1) Bit-depth
% ============================================================
\nextgroupplot[
    xlabel={Bit Depth},
    ylabel={Defense Success (\%)},
    ymin=0, ymax=25,
    ytick={0,5,10,15,20,25},
    xtick={1,2,3,4,5,6},
    xticklabels={2,3,4,5,6,7},
    xmin=0.2, xmax=6.8,
]
\addplot+[TwinLiteNetStyle]
coordinates {(1,1.5) (2,0.70) (3,11.97) (4,11.97) (5,7.75) (6,0.70)};

\addplot+[HybridNetStyle]
coordinates {(1,1.3) (2,1.6) (3,1.9) (4,1.47) (5,11.76) (6,2.94)};

\addplot+[CLRerNetStyle]
coordinates {(1,15.79) (2,8.33) (3,18.42) (4,3.51) (5,1.4) (6,0.44)};

\node[
    anchor=north,
    font=\footnotesize\bfseries,
    fill=white,
    fill opacity=0.7,
    text opacity=1,
    inner sep=1pt
] at (rel axis cs:0.5,0.98) {Bit-depth};

% ============================================================
% 2) Gaussian Noise
% ============================================================
\nextgroupplot[
    xlabel={Noise Std},
    ymin=0, ymax=5,
    ytick={0,1,2,3,4,5},
    xtick={1,2,3},
    xticklabels={0.02,0.05,0.1},
    xmin=0.3, xmax=3.7,
]
\addplot+[TwinLiteNetStyle]
coordinates {(1,0.70) (2,1.41) (3,0.70)};

\addplot+[HybridNetStyle]
coordinates {(1,1.47) (2,1.7) (3,1.2)};

\addplot+[CLRerNetStyle]
coordinates {(1,1.32) (2,0.44) (3,0.88)};

\node[
    anchor=north,
    font=\footnotesize\bfseries,
    fill=white,
    fill opacity=0.7,
    text opacity=1,
    inner sep=1pt
] at (rel axis cs:0.5,0.98) {Gaussian Noise};

% ============================================================
% 3) JPEG Compression
% ============================================================
\nextgroupplot[
    xlabel={Quality Factor},
    ymin=0, ymax=50,
    ytick={0,10,20,30,40,50},
    xtick={1,2,3,4,5,6},
    xticklabels={1,10,20,30,40,50},
    xmin=0.2, xmax=6.8,
]
\addplot+[TwinLiteNetStyle]
coordinates {(1,39.27) (2,16.20) (3,2.11) (4,9.86) (5,2.11) (6,4.93)};

\addplot+[HybridNetStyle]
coordinates {(1,43.10) (2,38.82) (3,23.53) (4,14.12) (5,4.41) (6,4.41)};

\addplot+[CLRerNetStyle]
coordinates {(1,44.56) (2,37.93) (3,28.86) (4,24.54) (5,12.39) (6,7.89)};

\node[
    anchor=north,
    font=\footnotesize\bfseries,
    fill=white,
    fill opacity=0.7,
    text opacity=1,
    inner sep=1pt
] at (rel axis cs:0.5,0.98) {JPEG Compression};

% ============================================================
% 4) Median Blur
% ============================================================
\nextgroupplot[
    xlabel={Kernel Size},
    ymin=0, ymax=5,
    ytick={0,1,2,3,4,5},
    xtick={1,2,3},
    xticklabels={10,15,20},
    xmin=0.3, xmax=3.7,
]
\addplot+[TwinLiteNetStyle]
coordinates {(1,1.8) (2,0.70) (3,2.82)};

\addplot+[HybridNetStyle]
coordinates {(1,1.1) (2,1.9) (3,2.94)};

\addplot+[CLRerNetStyle]
coordinates {(1,1.2) (2,1.6) (3,1.4)};

\node[
    anchor=north,
    font=\footnotesize\bfseries,
    fill=white,
    fill opacity=0.7,
    text opacity=1,
    inner sep=1pt
] at (rel axis cs:0.5,0.98) {Median Blur};

% ============================================================
% 5) Kalman Tracking
% ============================================================
\nextgroupplot[
    xlabel={Temporal Tracking},
    ymin=0, ymax=75,
    ytick={0,15,30,45,60,75},
    xtick={1},
    xticklabels={Kalman},
    xmin=0.4, xmax=1.6,
]
\addplot+[TwinLiteNetStyle]
coordinates {(1,68.0)};

\addplot+[HybridNetStyle]
coordinates {(1,63.3)};

\addplot+[CLRerNetStyle]
coordinates {(1,58.5)};

\node[
    anchor=north,
    font=\footnotesize\bfseries,
    fill=white,
    fill opacity=0.7,
    text opacity=1,
    inner sep=1pt
] at (rel axis cs:0.5,0.98) {Kalman Tracking};

\end{groupplot}

% ============================================================
% LEGEND
% ============================================================
\node[
    anchor=south,
    draw=black,
    font=\scriptsize
] at (current bounding box.north) {%
    \begin{tabular}{@{}l l l@{}}
        \tikz[baseline=-0.2ex]
        \fill[olive!80!black] (0,0) rectangle (0.35,0.15);
        ~TwinLiteNet
        &
        \tikz[baseline=-0.2ex]
        \fill[blue!80!black] (0,0) rectangle (0.35,0.15);
        ~HybridNets
        &
        \tikz[baseline=-0.2ex]
        \fill[red!80!black] (0,0) rectangle (0.35,0.15);
        ~CLRerNet
    \end{tabular}
};

\end{tikzpicture}%
}

% \vspace{1mm}

\caption{
Defense success rate (\%) against the \NSAttack on TwinLiteNet, HybridNets, and CLRerNet across four input pre-processing defenses and Kalman-filtered temporal tracking.
}

\vspace{-4mm}
\label{fig:defense-barplots}

\end{figure*}

We report the missed detection rate (MDR) as the percentage of attack trials in which participants failed to identify the presence of the \NSAttack during the relevant time interval. The MDR was 85\% under wet-road conditions and 80\% under clear-day conditions, indicating that most participants did not notice the attack. For the control videos without the attack, we separately report the no-alert rate, i.e., the percentage of trials in which participants did not flag any abnormality. Under clear conditions without the attack, the no-alert rate was 86\%. The similarity between the no-alert rate in normal conditions and the missed detection rate under attack conditions suggests that \NSes are difficult for human observers to distinguish from ordinary illumination artifacts. These findings highlight the stealthiness of the \NSAttack in SAE Level 1--2 driving scenarios, where human drivers are expected to supervise the vehicle and intervene when necessary. Here, stealthiness refers only to the visual conspicuousness of the projected \NS patterns to drivers; it does not imply that the physical setup is hidden or that the attacker remains unidentified.

\section{Countermeasures and Limitations}
\label{sec:9-Limitation_and_Defense_Discussion}

\noindent \textbf{Defenses.}
We evaluate whether input pre-processing and temporal defenses can mitigate the \NSAttack across different LD architectures. We focus on defenses that can be inserted before an existing LD pipeline without retraining the detector, also Kalman-filtered multi-frame lane tracking~\cite{kalman1960new}. Specifically, we consider four commonly used transformations: (i) JPEG compression with different quality factors, (ii) per-channel bit-depth reduction with different bit-widths, (iii) additive Gaussian noise with different standard deviations, and (iv) median blur with different kernel sizes.

For evaluation, we use only the frames from Section~\ref{sec:6-Misdetection_Evaluation} in which the \NSAttack successfully induces a spurious lane marking for a given LD model. For each such frame, we apply a pre-processing defense, re-run the corresponding LD algorithm on the transformed driver-view image, and mark the defense as \emph{successful} if the \NS region no longer produces a false lane marking. Figure~\ref{fig:defense-barplots} reports the resulting defense success rate, defined as the fraction of previously successful \NS samples neutralized by the defense.

Bit-depth reduction provides only limited protection. For example, using 4--5 bits per channel yields defense success rates of approximately $10$--$12\%$ for TwinLiteNet and HybridNets, and up to roughly $18\%$ for CLRerNet, while other bit-widths perform even worse. Gaussian noise and median blur are largely ineffective in our setting: across all tested parameter values, their defense success rates remain close to zero for all three models. This indicates that most \NS instances continue to be interpreted as plausible lane markings even after these perturbations.

JPEG compression is the most effective among the tested pre-processing defenses, but it remains insufficient. Very aggressive compression with quality factor $Q=1$ neutralizes roughly $39$--$45\%$ of successful \NSAttack instances across the three LD models. Intermediate qualities, such as $Q=10$ or $Q=20$, still provide non-trivial but lower protection. However, even in the best case, more than half of the \NS instances remain effective after compression. Moreover, such low-quality JPEG settings visibly degrade the image and would likely harm benign lane-keeping performance. Higher-quality configurations that are more suitable for normal driving, such as $Q \geq 30$, provide only modest protection, with defense success rates typically below $25\%$. Kalman-filtered temporal tracking provides stronger protection, preventing \NS-induced false-lane acceptance in 58.5--68.0\% of cases across the three LD models.

\noindent \textbf{Limitations.} First, we focused on L1/L2 AVs, whereas L3+ AVs utilize High Definition (HD) Maps, which provide precise road attributes such as curvature, slope, road width, lane width, and speed limits~\cite{liu2020high}. This prior knowledge reduces their reliance on LD algorithms, making safety violations less likely despite some misclassification by the \NSAttack. Second, while we conducted a real-world experiment to validate the attack, it posed significant challenges. Safety concerns, permission restrictions, and the need to create a controlled but realistic environment limited the scope of the experiment. We replicated a realistic road setup with standard lane dimensions, and shadow calculations to approximate real-world conditions under these constraints. However, achieving fully realistic and scalable scenarios in controlled environments remains challenging. 

In addition to real-world testing, we relied on advanced simulators like CARLA, which are widely recognized in both industry and academia for replicating real-world conditions safely and effectively~\cite{norden2019efficient,scanlon2021waymo,dosovitskiy2017carla,song2023discovering}. A miniature-scale testbed also provided valuable insights into the physical feasibility of the attack in controlled settings. CARLA and Comma 3X use their default automatic-exposure controls; we do not conduct a controlled exposure-parameter sweep, so sensitivity to different exposure policies remains unquantified. Finally, we do not explore sensor-fusion or training-time defenses such as adversarial training~\cite{goodfellow2014explaining} with \NS patterns; sensor fusion remains future work beyond our camera-based SAE Level 1--2 scope.

\section{Conclusion}
\label{sec:10-Conclusion}
We introduced the \NegativeShadow attack, a stealthy physical-world attack that exploits intensity bias in LD algorithms. The \NSAttack uses sunlight filtered through shaped openings to create bright lane-like patterns within shadowed regions. By parameterizing the attack with length, width, lateral distance, and angle, \NSes can mislead LD systems into interpreting these patterns as genuine lane markings. The attack was evaluated through LD misdetection analysis, software-in-the-loop simulations, a miniature road testbed, and a real-world-inspired field experiment. The results demonstrate that \NSes can induce false lane detections across diverse LD architectures and lead to safety-critical effects. A human-subject study further confirmed the stealthiness of the \NSAttack, while input-preprocessing defenses provided only limited protection.

\section*{Ethical Considerations}
\label{sec:ethical_considerations}
This research was conducted with careful consideration of ethical and safety implications. All human-subject experiments were approved by our university's IRB, ensuring informed consent, participant safety, and data anonymization. Because the attack targets safety-critical lane-keeping behavior, we avoided uncontrolled public-road testing. CARLA/Openpilot simulations and isolated physical experiments allowed us to evaluate hazardous scenarios under repeatable conditions without risk to pedestrians, surrounding traffic, or public infrastructure. Real-world validation was conducted only in a controlled private environment, and no public infrastructure was altered.

\section*{Acknowledgments}
We thank our anonymous reviewers. This work has been partially supported by the National Science Foundation (NSF) under grant CNS-2443252. Any findings, conclusions, and recommendations expressed in this paper are those of the authors and do not necessarily reflect the views of the NSF. This work was also partially supported by a gift from the BMW Group, as well as the Clemson University Creative Inquiry program. Assistance from undergraduate researcher Grace Johnson is gratefully acknowledged.

\bibliographystyle{IEEEtran}
\bibliography{ref}

\clearpage

% Multiple appendices in IEEEtran
\appendices

\numberwithin{equation}{section}

\section{Solar Position and Shadow Geometry Calculations}
\label{sec:11-Appendix_A}
\subsection{Solar Position Calculation}
\label{appendix:solar_angles}

For completeness, we summarize the standard solar-position equations used to obtain the solar altitude \(ALT\) and azimuth \(AZ\) from the calendar date, time, and observer location~\cite{shadow_calculator}. Let \(\phi\) and \(\lambda\) denote the observer's latitude and longitude, respectively, and let \(JD\) be the Julian date corresponding to the desired UTC time. We first compute the day number \(d\), which measures elapsed time from a fixed reference epoch:
\begin{equation}\label{eq:day_number_appendix}
d = JD - 2451545.
\end{equation}
The solar mean anomaly \(M\) (in degrees) is given by
\begin{equation}\label{eq:mean_anomaly_appendix}
M = 357.5291^\circ + 0.98560028^\circ \cdot d.
\end{equation}
We then apply the equation of the center \(C\) to correct for the eccentricity of Earth's orbit:
\begin{equation}\label{eq:center_equation_appendix}
C = 1.9148^\circ \sin M + 0.02^\circ \sin(2M) + 0.0003^\circ \sin(3M),
\end{equation}
and obtain the ecliptic longitude of the Sun,
\begin{equation}\label{eq:ecliptic_longitude_appendix}
L = M + C + 102.9372^\circ + 180^\circ.
\end{equation}

We convert from ecliptic to equatorial coordinates using the obliquity of the ecliptic \(\epsilon \approx 23.4397^\circ\). Writing all angles in radians for the trigonometric functions, we define
\begin{equation}\label{eq:equatorial_conversion_appendix}
\begin{aligned}
x_{\mathrm{equat}} &= \cos L, \\
y_{\mathrm{equat}} &= \cos \epsilon \,\sin L, \\
z_{\mathrm{equat}} &= \sin \epsilon \,\sin L.
\end{aligned}
\end{equation}
From these, the right ascension \(RA\) and declination \(\mathrm{Decl}\) follow as
\begin{equation}\label{eq:ra_decl_appendix}
RA = \arctan2(y_{\mathrm{equat}}, x_{\mathrm{equat}}), \qquad
\mathrm{Decl} = \arcsin(z_{\mathrm{equat}}).
\end{equation}

To account for Earth's rotation, we compute the local hour angle \(H\). Let \(\mathrm{GST}\) denote the Greenwich sidereal time corresponding to \(JD\), obtained from standard expressions (e.g.,~\cite{shadow_calculator}). The local sidereal time at longitude \(\lambda\) is
\begin{equation}\label{eq:lst_appendix}
LST = \mathrm{GST} + \lambda,
\end{equation}
and the hour angle is
\begin{equation}\label{eq:hour_angle_appendix}
H = LST - RA.
\end{equation}

Finally, the solar azimuth \(AZ\) and altitude \(ALT\) at the observer's location \((\phi,\lambda)\) are given by
\begin{equation}\label{eq:azimuth_altitude_appendix}
\begin{aligned}
AZ &= \arctan2~\!\bigl(\sin H,\; \cos H \,\sin \phi - \tan \mathrm{Decl} \,\cos \phi\bigr) \\ + 180^\circ, \\
ALT &= \arcsin~\!\bigl(\sin \phi \,\sin \mathrm{Decl} + \cos \phi \,\cos \mathrm{Decl} \,\cos H\bigr).
\end{aligned}
\end{equation}

In the main text (Section~\ref{sec:5_2_Shadow_Calculation}), we treat \(ALT\) and \(AZ\) as the outputs of this standard pipeline and use them directly in the geometric shadow-length formula.

\subsection{Solar Feasibility Analysis}

We conduct a five-year analysis using 5-minute NSRDB solar-irradiance data~\cite{sengupta2018nsrdb} from 2021--2025 at five U.S. locations: Phoenix (AZ), Las Vegas (NV), Albuquerque (NM), Seattle (WA), and Pittsburgh (PA). We vary road/canopy orientation in $15^\circ$ increments and consider mounting heights of 5 and 10\,m. Every daylight interval is included, and an interval is considered available when $SL \geq 25$\,m and $DNI \geq 400$\,W/m$^2$; cloudy, low-irradiance, and geometrically unsuitable intervals are counted as unavailable. The annual median across orientations ranges from 2.22--4.14\% of daylight intervals at 5\,m and 9.38--17.25\% at 10\,m; seasonal medians range from 1.53--5.37\% and 7.41--19.45\%, respectively. These percentages quantify recurring solar-geometric windows conditional on suitable adjacent property or infrastructure and do not estimate site prevalence or end-to-end attack success.

\section{Ablation Details}
\label{sec:15-Appendix_E}
\subsection{Illumination-Aware Ablation}

\noindent \textbf{Ablation Protocol.}
When ablating a given parameter $X \in \{W,L,D,\beta\}$, we first construct a pool of approximately 100 \NS configurations by sampling the other three parameters uniformly within narrow tolerance bands around the base configuration,
$(W_0,L_0,D_0,\beta_0) = (12\,\mathrm{cm},30\,\mathrm{m},10\,\mathrm{cm},8^\circ)$.
Specifically, we sample within bands such as
$|W - W_0| \leq 5\,\mathrm{cm}$,
$|L - L_0| \leq 5\,\mathrm{m}$,
$|D - D_0| \leq 5\,\mathrm{cm}$, and
$|\beta - \beta_0| \leq 4^\circ$.
This pool of unablated triples is \emph{kept identical} across all tested levels of $X$. For each level $x \in \hat{X}$, we replace the corresponding coordinate in every triple with $X=x$, cast the resulting \NS with the desired luminance offset in BEV, warp it to the driver view, and evaluate all three LD algorithms. Thus, in each subplot of \autoref{fig:ablation_all_one_row}, every ring, corresponding to a fixed level of $X$, and every illumination sector, corresponding to a luminance offset $\Delta L^*$, reports a success rate computed over the same randomized configurations of the remaining parameters. This design allows different values in $\hat{W}$, $\hat{L}$, $\hat{D}$, and $\hat{\beta}$ to be compared under the same background geometry, so changes in \NS success rate can be attributed primarily to the chosen level of the ablated parameter under varying illumination.

\noindent \textbf{Illumination Calibration Details.}
For a set of representative daytime CARLA weather presets, we render a scene with an \NS, sweep $\Delta L^*$ in our rendering pipeline, and select the value whose \NS--road contrast most closely matches the corresponding CARLA frame. Collecting these best-match offsets across all selected presets yields six representative illumination settings,
$\Delta L^* \in \{5,10,15,20,25,30\}$.
These values are then used consistently in the illumination-aware ablation in \autoref{fig:ablation_all_one_row}.

\noindent \textbf{Statistical Interpretation.}
For each ablation setting, we estimate the probability that a randomly sampled \NS within the local tolerance tube around $(W_0,L_0,D_0,\beta_0)$ fools a given detector. Fix a parameter $X \in \{W,L,D,\beta\}$, a level $x \in \hat{X}$, and a luminance offset $\Delta L^*$. As described above, the remaining three parameters are sampled uniformly within narrow bands around the base configuration, and the same pool of approximately $n=100$ geometric triples is reused for all levels of $X$ and all $\Delta L^*$. For each triple, we cast the corresponding \NS, run the detector, and record a binary outcome
$y_j(x,\Delta L^*) \in \{0,1\}$,
indicating whether the \NS is misclassified as a lane marking. We model these binary outcomes as Bernoulli trials and summarize each $(x,\Delta L^*)$ cell by its empirical success rate:
{%
  \setlength{\abovedisplayskip}{4pt}
  \setlength{\belowdisplayskip}{4pt}
  \begin{equation}
    \hat{p}(x,\Delta L^*)
    =
    \frac{1}{n}\sum_{j=1}^{n} y_j(x,\Delta L^*).
  \end{equation}
}

To obtain a discrete, visually interpretable notion of risk, we map $\hat{p}(x,\Delta L^*)$ into three categories: high-risk if $\hat{p}(x,\Delta L^*) \geq 0.7$, low-risk if $\hat{p}(x,\Delta L^*) \leq 0.3$, and intermediate otherwise. With $n \approx 100$ trials per cell, these thresholds separate configurations where the attack succeeds on a clear majority of realizations from those where it fails on most realizations. In the color-coded summaries, high-risk cells are shown in green, low-risk cells in red, and intermediate cells in yellow; nighttime intervals, where we do not simulate any \NS, are not shown.

\subsection{Mechanism Ablation}
\label{sec:mechanism_ablation}
We conduct an ablation on 2,000 configurations at $\Delta L^*=30$. The attack succeeds on 403, 211, and 263 configurations against TwinLiteNet, HybridNets, and CLRerNet, respectively. For each successful configuration, we reduce the local luminance contrast inside the mapped \NS region using Gaussian-smoothed local illumination while preserving its geometry, position, perspective, detector settings, and all pixels outside the region; as a spatial control, we apply the same adjustment to a similarly sized non-\NS road region. This eliminates false-lane detection in 99.3\%, 99.1\%, and 84.4\% of successful attacks, versus 18.1\%, 4.3\%, and 1.5\% for the control intervention (exact McNemar test \cite{fagerland2013mcnemar}, $p<10^{-4}$). These results show that local bright-within-shadow contrast is a key factor in \NS success, consistent with the proposed intensity-bias explanation, but do not isolate a specific shadow-removal implementation.

\section{Safety Violation: Reaction Time}
\label{sec:12-Appendix_B}
The average driver reaction time with ALC systems may exceed 2.5 seconds~\cite{sato2021dirty}, so the driver must take over within 2.5 seconds to counter the attack effectively. For each scenario, the vehicle was driven at different speeds, and the reaction time was calculated based on the distance between the deviation at \NS detection and the goal defined in Section~\ref{sec:7_1_Software-in-the-Loop}. \autoref{tab:reaction_time} shows the time of attack occurrence for each scenario described in Section~\ref{sec:7_1_Software-in-the-Loop}. Reaction times within 2.5 seconds are indicated by \textcolor{green!70!black}{\rule{0.7em}{0.7em}}, while reaction times longer than 2.5 seconds are indicated by \textcolor{red!70!black}{\rule{0.7em}{0.7em}}.

For each speed, the patterns in \autoref{tab:reaction_time} show where the driver has enough time to counter the attack (green) and where the available reaction time is no longer sufficient (red). As speed increases, the same \NS length leaves less time for the driver to react, and longer \NSes within the 25--35\,m range further reduce this margin. By combining the outcomes of \autoref{tab:analysis_table} and \autoref{tab:reaction_time} using an AND approach, \autoref{tab:reaction_time_and_success} shows the scenarios where the attack was successful and the driver had sufficient time to react.

\begin{table}[h]
\tiny
\centering
\caption{Evaluation of driver reaction times to ALC system attacks at various speeds and shadow lengths} 
\label{tab:reaction_time}
\resizebox{\columnwidth}{!}{
\begin{tabular}{cccccc}
\toprule
\multicolumn{6}{c}{ \textbf{Speed (mph)}} \\
\cline{2-6}
\textbf{Length (m)} & \textbf{10} & \textbf{15} & \textbf{20} & \textbf{35} & \textbf{60} \\
\hline
25 &
\textcolor{green!70!black}{\rule{0.7em}{0.7em}}\textcolor{red!70!black}{\rule{0.7em}{0.7em}}\textcolor{red!70!black}{\rule{0.7em}{0.7em}} &
\textcolor{green!70!black}{\rule{0.7em}{0.7em}}\textcolor{green!70!black}{\rule{0.7em}{0.7em}}\textcolor{red!70!black}{\rule{0.7em}{0.7em}} &
\textcolor{green!70!black}{\rule{0.7em}{0.7em}}\textcolor{green!70!black}{\rule{0.7em}{0.7em}}\textcolor{red!70!black}{\rule{0.7em}{0.7em}} &
\textcolor{green!70!black}{\rule{0.7em}{0.7em}}\textcolor{green!70!black}{\rule{0.7em}{0.7em}}\textcolor{green!70!black}{\rule{0.7em}{0.7em}} &
\textcolor{green!70!black}{\rule{0.7em}{0.7em}}\textcolor{red!70!black}{\rule{0.7em}{0.7em}}\textcolor{green!70!black}{\rule{0.7em}{0.7em}} \\
30 &
\textcolor{green!70!black}{\rule{0.7em}{0.7em}}\textcolor{red!70!black}{\rule{0.7em}{0.7em}}\textcolor{red!70!black}{\rule{0.7em}{0.7em}} &
\textcolor{green!70!black}{\rule{0.7em}{0.7em}}\textcolor{green!70!black}{\rule{0.7em}{0.7em}}\textcolor{red!70!black}{\rule{0.7em}{0.7em}} &
\textcolor{green!70!black}{\rule{0.7em}{0.7em}}\textcolor{green!70!black}{\rule{0.7em}{0.7em}}\textcolor{green!70!black}{\rule{0.7em}{0.7em}} &
\textcolor{green!70!black}{\rule{0.7em}{0.7em}}\textcolor{green!70!black}{\rule{0.7em}{0.7em}}\textcolor{green!70!black}{\rule{0.7em}{0.7em}} &
\textcolor{green!70!black}{\rule{0.7em}{0.7em}}\textcolor{red!70!black}{\rule{0.7em}{0.7em}}\textcolor{green!70!black}{\rule{0.7em}{0.7em}} \\
35 &
\textcolor{green!70!black}{\rule{0.7em}{0.7em}}\textcolor{red!70!black}{\rule{0.7em}{0.7em}}\textcolor{red!70!black}{\rule{0.7em}{0.7em}} &
\textcolor{green!70!black}{\rule{0.7em}{0.7em}}\textcolor{green!70!black}{\rule{0.7em}{0.7em}}\textcolor{green!70!black}{\rule{0.7em}{0.7em}} &
\textcolor{green!70!black}{\rule{0.7em}{0.7em}}\textcolor{green!70!black}{\rule{0.7em}{0.7em}}\textcolor{green!70!black}{\rule{0.7em}{0.7em}} &
\textcolor{green!70!black}{\rule{0.7em}{0.7em}}\textcolor{green!70!black}{\rule{0.7em}{0.7em}}\textcolor{green!70!black}{\rule{0.7em}{0.7em}} &
\textcolor{green!70!black}{\rule{0.7em}{0.7em}}\textcolor{red!70!black}{\rule{0.7em}{0.7em}}\textcolor{green!70!black}{\rule{0.7em}{0.7em}} \\
\bottomrule
\end{tabular}}
\end{table}

\begin{table}[h]
\tiny
\centering 
\caption{Comprehensive overview of scenarios where the attack succeeds and the driver has sufficient time to react (green) versus all other cases (red).}
\label{tab:reaction_time_and_success}
\resizebox{\columnwidth}{!}{
\begin{tabular}{cccccc}
\toprule
\multicolumn{6}{c}{\textbf{Speed (mph)}} \\
\cline{2-6}
\textbf{Length (m)} & \textbf{10} & \textbf{15} & \textbf{20} & \textbf{35} & \textbf{60} \\
\hline
\textbf{25} &
\textcolor{green!70!black}{\rule{0.7em}{0.7em}}\textcolor{red!70!black}{\rule{0.7em}{0.7em}}\textcolor{red!70!black}{\rule{0.7em}{0.7em}} &
\textcolor{green!70!black}{\rule{0.7em}{0.7em}}\textcolor{green!70!black}{\rule{0.7em}{0.7em}}\textcolor{red!70!black}{\rule{0.7em}{0.7em}} &
\textcolor{green!70!black}{\rule{0.7em}{0.7em}}\textcolor{green!70!black}{\rule{0.7em}{0.7em}}\textcolor{red!70!black}{\rule{0.7em}{0.7em}} &
\textcolor{green!70!black}{\rule{0.7em}{0.7em}}\textcolor{green!70!black}{\rule{0.7em}{0.7em}}\textcolor{green!70!black}{\rule{0.7em}{0.7em}} &
\textcolor{green!70!black}{\rule{0.7em}{0.7em}}\textcolor{red!70!black}{\rule{0.7em}{0.7em}}\textcolor{green!70!black}{\rule{0.7em}{0.7em}} \\
\textbf{30} &
\textcolor{green!70!black}{\rule{0.7em}{0.7em}}\textcolor{red!70!black}{\rule{0.7em}{0.7em}}\textcolor{red!70!black}{\rule{0.7em}{0.7em}} &
\textcolor{green!70!black}{\rule{0.7em}{0.7em}}\textcolor{green!70!black}{\rule{0.7em}{0.7em}}\textcolor{red!70!black}{\rule{0.7em}{0.7em}} &
\textcolor{green!70!black}{\rule{0.7em}{0.7em}}\textcolor{green!70!black}{\rule{0.7em}{0.7em}}\textcolor{green!70!black}{\rule{0.7em}{0.7em}} &
\textcolor{green!70!black}{\rule{0.7em}{0.7em}}\textcolor{green!70!black}{\rule{0.7em}{0.7em}}\textcolor{green!70!black}{\rule{0.7em}{0.7em}} &
\textcolor{green!70!black}{\rule{0.7em}{0.7em}}\textcolor{red!70!black}{\rule{0.7em}{0.7em}}\textcolor{red!70!black}{\rule{0.7em}{0.7em}} \\
\textbf{35} &
\textcolor{green!70!black}{\rule{0.7em}{0.7em}}\textcolor{red!70!black}{\rule{0.7em}{0.7em}}\textcolor{red!70!black}{\rule{0.7em}{0.7em}} &
\textcolor{green!70!black}{\rule{0.7em}{0.7em}}\textcolor{green!70!black}{\rule{0.7em}{0.7em}}\textcolor{green!70!black}{\rule{0.7em}{0.7em}} &
\textcolor{green!70!black}{\rule{0.7em}{0.7em}}\textcolor{green!70!black}{\rule{0.7em}{0.7em}}\textcolor{green!70!black}{\rule{0.7em}{0.7em}} &
\textcolor{green!70!black}{\rule{0.7em}{0.7em}}\textcolor{green!70!black}{\rule{0.7em}{0.7em}}\textcolor{green!70!black}{\rule{0.7em}{0.7em}} &
\textcolor{green!70!black}{\rule{0.7em}{0.7em}}\textcolor{red!70!black}{\rule{0.7em}{0.7em}}\textcolor{green!70!black}{\rule{0.7em}{0.7em}} \\
\bottomrule
\end{tabular}}
\vspace{2pt}

{\scriptsize
\textcolor{green!70!black}{\rule{0.6em}{0.6em}} = \textbf{scenario success AND reaction time $\leq 2.5$s},\quad
\textcolor{red!70!black}{\rule{0.6em}{0.6em}} = otherwise.
}
\end{table}

\section{Miniature Testbed Construction Details}
\label{sec:14-Appendix_D}
We constructed the miniature road model to approximate a full-scale reference road of 50\,m in length. A typical U.S. street is 360\,cm wide~\cite{FHWA2023DesigningRoadDiet}, with lane markings approximately 16\,cm wide~\cite{civilsir2024,FHWA2023PavementMarking}. Our miniature road is 45\,cm wide and 165\,cm long. To scale the lane markings, we used the ratio between the lane-marking width and the road width in the full-scale setting, i.e., \(16/360\). Applying this ratio to the 45\,cm-wide miniature road gives a scaled lane-marking width of approximately \(2\,\mathrm{cm}\).

Because the line markings on the reference road are dashed, we also scaled the dash pattern. The typical pattern of 3\,m-long markings followed by 9\,m gaps was proportionally reduced using the ratio between the miniature road length, 165\,cm, and the reference road length, 5000\,cm. This results in lane markings approximately 9.9\,cm long, separated by 29.7\,cm gaps.

A flat wooden panel was placed parallel to the road surface with a magnet positioned on top of the panel. A second magnet was attached below the panel to one end of a pipe, while the Comma 3X device running Openpilot v0.9.5 was connected to the other end. Moving the upper magnet manually caused the lower magnet, pipe, and Comma 3X to move through magnetic attraction.

To cast two parallel \NSes over the miniature road, we mounted a canopy with two parallel openings, each 0.31\,cm wide and 42\,cm long. The light source was fixed at an angle of \(46.7^\circ\) relative to the miniature road. The canopy was placed 83\,cm horizontally and 61\,cm vertically from the road model, producing \NS parameters within the successful parameter ranges: \(D \approx 9\,\mathrm{cm}\), \(\beta \approx 5^\circ\), \(W \approx 2.2\,\mathrm{cm}\), and \(L \approx 64\,\mathrm{cm}\).

\section{Human Study Protocol and Participant Demographics}
\label{sec:13-Appendix_C}
We provide additional details on the human-subject study protocol used to evaluate the stealthiness of the \NSAttack. Participants watched prerecorded CARLA videos generated under different weather conditions. To reduce response bias, participants were told that the study evaluated driving videos under different weather conditions in a simulated environment; the true purpose of the \NSAttack evaluation was revealed only after the experiment.

\begin{table}[h]
    \centering
    \renewcommand{\arraystretch}{1}
    \caption{Demographic information of participants}
    \label{table:Demographic}
    \resizebox{\columnwidth}{!}{%
        \begin{tabular}{lll}
            \hline
            \textbf{Questions} & \textbf{Options} & \textbf{Percentages} \\
            \hline
            \multirow{6}{*}{Age}
                & 18--19      & 5\%  \\
                & 20--21      & 17\% \\
                & 22--23      & 23\% \\
                & 24--25      & 25\% \\
                & 26--27      & 18\% \\
                & 28 or older & 12\% \\
            \hline
            \multirow{3}{*}{Gender}
                & Male   & 80\% \\
                & Female & 20\% \\
                & Other  & 0\%  \\
            \hline
            \multirow{3}{*}{Ethnicity}
                & Hispanic or Latino or Spanish Origin            & 0\% \\
                & Not Hispanic or Latino or Spanish Origin        & 97\% \\
                & I do not wish to provide this information       & 3\%  \\
            \hline
            \multirow{6}{*}{Race}
                & Asian                                      & 72\% \\
                & Black or African American                  & 8\%  \\
                & Native Hawaiian or Other Pacific Islander  & 0\%  \\
                & White                                      & 20\% \\
                & Hispanic/Latino                            & 0\%  \\
                & I do not wish to provide this information  & 0\%  \\
            \hline
            \multirow{2}{*}{Driver's License}
                & Yes & 100\% \\
                & No  & 0\%   \\
            \hline
            \multirow{2}{*}{AV Experience}
                & Somewhat & 50\% \\
                & Never    & 50\% \\
            \hline
        \end{tabular}%
    }
\end{table}

Before the experiment, participants completed a short demographic questionnaire. They were then instructed to watch each prerecorded AV driving video and press the space bar whenever they felt that something was abnormal in the simulation. Participants were also instructed to press the space bar a second time when they believed the abnormal situation had been resolved. This setup reflects SAE Level 1--2 driving systems, where the human driver is expected to supervise the vehicle and intervene when necessary.

To measure participant responses, we implemented a Python script that recorded keyboard inputs and video timestamps. When a participant pressed the space bar, the script recorded the corresponding timestamp in the video. A second press recorded the end timestamp of the perceived abnormal event. The script then automatically computed the duration of each response interval. These timestamps allowed us to compare participant responses with the time intervals during which the \NSes were present. Before conducting the official study, we ran a small pilot study to verify that the video playback, keyboard logging, and timestamp recording worked correctly. After validating the setup, we recruited 60 participants anonymously for the official study.

After the video trials, participants answered post-experiment questions to determine whether they had inferred the true purpose of the study before debriefing. Participants who indicated prior awareness of the study objective were excluded from the analysis. At the end of the study, participants were debriefed and informed of the actual purpose of the experiment. We report the demographic distribution of the 60 participants in \autoref{table:Demographic}.

\section*{LLM Usage Statement}
LLMs were used for editorial purposes (proofreading and grammar checking) in this manuscript. All outputs were inspected by the authors to ensure accuracy and originality.

\end{document}